\documentclass{emulateapj}

\begin{document}
\title{The Population of Helium-Merger Progenitors:  Observational Predictions}

\def\msun{{\rm ~M}_{\odot}}
\def\rsun{{\rm ~R}_{\odot}}
\def\myr{{\rm ~Myr}}
\def\mdot{\dot M}
\def\mpy{{\rm ~M}_{\odot} {\rm ~yr}^{-1}}

\author{Chris L. Fryer\altaffilmark{1,2,3},Krzysztof
  Belczynski\altaffilmark{4,5}, Edo Berger\altaffilmark{6}, Christina
  Th\"one\altaffilmark{7}, Carola Ellinger\altaffilmark{8}, Tomasz
  Bulik\altaffilmark{4}}

\altaffiltext{1}{CCS Division, Los Alamos National Laboratory, 
Los Alamos, NM 87545, USA}
\altaffiltext{2}{Department of Physics, The University of Arizona,
Tucson, AZ 85721, USA} 
\altaffiltext{3}{Department of Physics and Astronomy, The University of New Mexico,
Albuquerque, NM 87131, USA} 
\altaffiltext{4}{Astronomical Observatory, University of Warsaw, Al. Ujazdowskie 4, 00-478 Warsaw, Poland}
\altaffiltext{5}{Center for Gravitational Wave Astronomy, University of Texas at Brownsville, TX 78520, USA}
\altaffiltext{6}{Harvard-Smithosonian Center for Astrophysics, 60 Garden Street, Cambridge, MA 02138, USA}
\altaffiltext{7}{IAA-CSIC, Glorieta de la Astronom\'ia s/n, 18008 Granada, Spain}
\altaffiltext{8}{Department of Physics, University of Texas at Arlington, 502 Yates Street, Box 19059, Arlington, TX 76019, USA}

\begin{abstract}

The helium-merger gamma-ray burst progenitor is produced by the rapid
accretion onto a compact remnant (neutron star or black hole) when it
undergoes a common envelope inspiral with its companion's helium
core. This merger phase produces a very distinct environment around
these outbursts and recent observations suggest that, in some cases,
we are detecting the signatures of the past merger in the GRB
afterglow. These observations allow us, for the first time, to study
the specific features of the helium merger progenitor.  In this paper,
we couple population synthesis calculations to our current
understanding of gamma-ray burst engines and common envelope evolution
to make observational predictions for the helium-merger gamma-ray
burst population.  Many mergers do not produce GRB outbursts and 
we discuss the implications of these mergers with the broader 
population of astrophysical transients.  

\end{abstract}


\section{Introduction}

The black hole accretion disk gamma-ray burst (GRB) engine has become
standard for long- and short-duration GRBs alike~\citep{Popham99}.
For long-duration bursts, the most commonly-invoked progenitor is the
collapse of a rapidly rotating massive star down to a black hole.  But
achieving the high angular momentum needed to produce a disk around
the newly formed black hole has proved problematic and a number of
progenitors have been invoked to provide these high angular momentum
profiles~\citep{FW98,Fry99a,yoon2005,woosley2006,belczynski2007,ivanova2003,portegieszwart2005,vdheuvel2007,fryer2007a,podsiadlowski2010}.

One such model invokes the merger of a compact remnant with its
companion.  As the compact remnant spirals into the center of the
helium core of an evolved companion, the accretion rate can exceed
0.01\,M$_\odot$\,s$^{-1}$, enough to power a strong gamma-ray
burst~\citep{FW98}.  The basic idea behind this model is that as the
compact remnant spirals into its companion, the orbital energy lost
both spins up the helium core and ejects the hydrogen envelope.  If
the compact remnant is a neutron star, the rapid accretion in the
inspiral is likely to cause it to collapse to a black hole.  In this
manner, the helium merger model provides a natural way to ensure
enough angular momentum to form a disk around a central black hole.

Simulations by Zhang \& Fryer (2001) showed that the inspiral would
produce both the high angular momenta and accretion rates needed to
create a black hole accretion disk engine.  Although there is some
concern that helium mergers produce disks with too much angular
momentum~\citep{DiM02}, current analyses find this angular momentum is
appropriate for accretion disk engines\citep{Kom09,Bar11}.  The
helium-merger model predicts a feature that may be observable in the
resultant explosion: the existence of a shell (or torus) of merger-ejecta 
surrounding the burst.

Initially, \cite{FW98} believed the helium merger model would be
difficult to distinguish from normal GRBs because the ejecta from the
common envelope phase would lie along the orbital plane (the same
plane as the disk) and there would be very little interaction between
the ejecta and the GRB jet.  This picture changed with the observation
of GRB121225, the so-called ``Christmas Burst''\citep{Tho11}.  This
burst exhibited blackbody emission arguing for a compact progenitor
($\sim 10^{11}$\,cm) followed by a second blackbody component with
characteristic radius of $\sim 10^{14}$\,cm.  Simulations of the jet
interactions with more realistic common envelope ejecta profiles (also
based on recent simulations) found that the helium merger model could
produce the thermal emission in this burst~\citep{Tho11}.  The X-ray
Flare 060218 (and perhaps a few other low-redshift bursts) may have
similar characteristics, suggesting that these helium merger events
may be identifiable after all.

To truly compare these systems to the observed data, we must
understand the entire population of compact merger events.  In this
paper, we undertake a series of population synthesis models studying
the characteristics of compact merger events.  Section~\ref{sec:population}
describes the population synthesis calculations and presents the basic
results of these calculations: rates and characteristics of merging
systems.  In section~\ref{sec:observations}, we analyze these results, 
making predictions for luminosities and ejecta positions for merger 
outbursts.  We conclude with a discussion of the potential observational 
outcomes of this black hole accretion disk progenitor.

\section{Neutron Star Merger Populations}
\label{sec:population}

\subsection{Population Synthesis Calculations}

To study the helium merger progenitor, we use the population synthesis
code {\tt StarTrack} to calculate the numbers and properties of
merging systems. The full description of the code can be found
in~\cite{belczynski2002,belczynski2008}.  The code utilizes a set of
stellar models (\cite{hurley2000}; slightly modified from its original
version) that allow for the evolution of stars at different
metallicities. The model for compact object formation adopted in the
code has been significantly revised~\citep{dominik2012}. During core
collapse, fallback and direct BH formation is now accounted
for~\citep{fryer2001} and the newly born objects receive natal
kicks~\citep{hobbs2005}. The formation of low mass NSs through
electron capture supernovae is also accounted for (e.g.,
\cite{podsiadlowski2004}). Binary interactions are treated in detail,
and the various physical processes have been calibrated using either
results of detailed evolutionary calculations (e.g.,
\cite{wellstein1999} for mass transfer sequences), or specific sets of
observations (e.g., \cite{levine2000} for tidal interactions). The
mass loss prescription for winds in massive stars, the stellar binding
energy used in common envelope evolution, and the compact remnant mass
formation prescription have all been updated~\citep{dominik2012}.

\subsection{Rates}
\label{sec:poprate}

In each calculation, we have evolved $N=2 \times 10^5$ binaries  with the primary 
mass range $M_1=5-150 \msun$ and secondary mass range $M_1=0.08-150 \msun$. 
The stars were evolved with solar metallicity ($Z=0.02$). 
Each binary system is initiated by four parameters, which are assumed to be
independent: the primary mass $M_1$ (the initially more massive component), 
the mass ratio $q={M_2 \over M_1}$, where $M_2$ is the mass of the secondary,
the semi-major axis $a$ of the orbit, and the orbital eccentricity $e$. 

For both, single stars and binary system primaries, we use the initial
mass function adopted from Kroupa \& Weidner (2003),
\begin{equation}
 \Psi(M) \propto \left\{ 
              \begin{array}{ll}
                {M}^{-1.3} & 0.08 \leq M < 0.5 \msun \\
                {M}^{-2.2} & 0.5 \leq M < 1.0 \msun \\
                {M}^{-\alpha_{\rm imf}} & 1.0 \leq M < 150 \msun \\
              \end{array}
            \right.
\label{init01}
\end{equation}
where parameter $\alpha_{\rm imf}=2.35-3.2$, with our standard choice of 2.7. 

We adopt a flat mass ratio distribution that is consistent with the recent
observational results (Kobulnicki, Fryer \& Kiminki 2006),
 \begin{equation}
 \Phi(q) = 1 
 \end{equation} 
 in the range $q=0-1$.  Given value of the primary mass
and the mass ratio, we obtain the mass of the secondary $M_2= q M_1$. 
 
The distribution of the initial binary separations is assumed to be 
flat in the logarithm (Abt 1983),
 \begin{equation}
 \Gamma(a) \propto {1\over a},
 \end{equation} 
 where $a$ ranges from a minimum value, such that the
primary fills at maximum 50\% of its Roche lobe at ZAMS, up to $10^5 \, 
{\rm R}_\odot$.

We adopt the thermal-equilibrium eccentricity distribution for initial
binaries,
 \begin{equation}
 \Xi(e) = 2e,  
 \end{equation} 
in the range $e = 0-1$ (e.g., Heggie 1975; Duquennoy \& Mayor 1991).

We have evolved only binaries massive enough to potentially produce at least one 
compact object and therefore produce a helium merger. If we extend the evolved 
population to include the entire mass spectrum  and if we include single stars as 
well (with binary fraction $f_{\rm bi}=0.5$), each calculation with 
$\alpha_{\rm imf}=-2.7$ would correspond to a stellar mass formed in a starburst 
mass of $4.5 \times 10^7 \msun$. 

\begin{deluxetable}{lccc}
\tablewidth{0pt} \tablecaption{Population Synthesis Models} \tablehead{
  \colhead{Model} & \colhead{$\lambda_{CE}$} & \colhead{Hertzprung Gap}  & \colhead{Metallicity} \\ 
  \colhead{} & \colhead{} & \colhead{Fate\tablenotemark{a}} & \colhead{}}

\startdata
Model1 & physical $\lambda$\tablenotemark{b} & CE formulae & solar \\
Model2 & 3$\times$physical $\lambda$ & CE forumulae & solar \\
Model3 & constant $\lambda=1$ & CE formulae & solar \\
Model4 & physical $\lambda$ & HG always merges & solar \\
Model5 & physical $\lambda$ & CE formulae & 10\% solar \\
Model6 & physical $\lambda$ & HG always merges & 10\%solar \\
\enddata

\tablenotetext{a}{Stars in the Hertzprung gap do not have strong
  entropy gradients dividing the helium core and hydrogen envelope.
  Without these sharp gradients, every CE phase will proceed through
  the merger with the helium core.  We choose two fates:  ignore this 
entropy argument (``CE formulae'') or include this effect and assume 
on Hertzprung gap CE scenarios lead to a merged system (``HG always merges'').  
For more details, see \cite{dominik2012}.}
\tablenotetext{b}{Physical $\lambda$ corresponds to the \cite{xu2010} used 
by \cite{dominik2012}.}

\label{table:popdesc}

\end{deluxetable}

Table~\ref{table:popresults} shows the total number of he-mergers
produced using $2\times10^5$ binaries.  The rates are divided into 3
categories: main-sequence merger and mergers with small, and large,
helium core masses.  These rates correspond to the rate per $4.5
\times 10^7 \msun$ of star forming mass.  The number in parantheses
show the number of the total systems produced because of neutron star
kicks. Although kicks produce many of the merging systems (30\%),
kick-induced mergers account for a smaller fraction ($\sim$10\%) of
low-mass helium core mergers and almost nobe ($<$1\%) of the massive helium
mergers believed to be GRBs.

\begin{deluxetable}{lccc}
\tablewidth{0pt} \tablecaption{Merger Rates (per $4.5\times10^7$\,M$_\odot$)} \tablehead{
  \colhead{Model} & \colhead{He-Merger} & \colhead{He-Merger Rate}  & \colhead{Main-Sequence} \\ 
  \colhead{} & \colhead{($M_{\rm He} < 4.0M_\odot$)} & \colhead{($M_{\rm He} > 4.0M_\odot$)} & \colhead{Merger Rate}}

\startdata
Model1 & 718(110)\tablenotemark{a} & 939(0) & 579(175) \\
Model2 & 507(35) & 406(0) & 1267(411) \\
Model3 & 967(94) & 227(3) & 1566(478) \\
Model4 & 929(24) & 1020(2) & 580(175) \\
Model5 & 1505(1) & 2019(9) & 796(249) \\
Model6 & 1790(1) & 3033(6) & 755(262) \\

\enddata

\tablenotetext{a}{The left number in each column denotes the total 
number of systems produced in each category.  The numbers in parantheses 
show the number of these systems formed due to neutron star kicks.}
\label{table:popresults}

\end{deluxetable}

These rates can be folded with any desired star formation history to
give actual time dependent helium merger rate in a given galaxy.
Table~\ref{table:popresults2} shows the merger rate per Myr in a
Milky-Way massed galaxy as well as the volumetric (Gpc$^{-3}$
yr$^{-1}$) rate.  If we wish to compare to GRBs, a rough estimate of
the rates should include a combination of solar and low metallicity
rates (e.g. 50\% each of Model 1 and Model 5 or 50\% each of Model 4
and Model 6).  The resultant rates predicted are 629, 857 Gpc$^{-3}$
yr$^{-1}$ for respective models assuming normal CE formulae for
Hertzsprung gap mergers versus immediate mergers for these Hertzsprung
gap mergers.

The Fryer et al. (1999) Galactic rate for their standard model is
35\,Myr$^{-1}$, very close to our rate of high-mass helium cores.  Such 
high rates mean that even if only 10\% of these systems produce GRB 
jets, they can explain a large fraction of the 
GRB rates.  The low-mass helium star merger and main-sequence merger 
rate is not significantly higher and, assuming these mergers produce 
peculiar supernovae, they can account for no more than 1\% of all 
core-collapse sueprnovae.

\begin{deluxetable}{lccc}
\tablewidth{0pt} \tablecaption{Merger Rates II (Myr$^{-1}$, Gpc$^{-3}$ yr$^{-1}$)} \tablehead{
  \colhead{Model} & \colhead{He-Merger} & \colhead{He-Merger Rate}  & \colhead{Main-Sequence} \\ 
  \colhead{} & \colhead{($M_{\rm He} < 4.0M_\odot$)} & \colhead{($M_{\rm He} > 4.0M_\odot$)} & \colhead{Merger Rate}}

\startdata
Model1 & 30.8(308.2)\tablenotemark{a} & 40.3(403.0) & 24.9(248.5) \\
Model2 & 22.0(220.3) & 17.6(176.4) & 55.0(550.5) \\
Model3 & 43.2(431.5) & 10.1(101.3) & 69.9(69.9) \\
Model4 & 39.9(399.2) & 43.8(438.3) & 24.9(249.3) \\
Model5 & 63.7(636.7) & 85.4(854.2) & 33.7(336.8) \\
Model6 & 75.2(752.4) & 127.5(1274.8) & 31.7(317.3) \\

\enddata

\tablenotetext{a}{The left number in each column denotes a rate per Milky-Way 
massed galaxy (Myr$^{-1}$).  The numbers in parantheses 
show volumetric rate (Gpc$^{-3}$ yr$^{-1}$).}
\label{table:popresults2}

\end{deluxetable}

Like long-duration GRBs, the delay time from star
formation to merged event is limited to the stellar evolution time (in
the he-merger case, it is the stellar evolution time of the companion
star).  For massive helium mergers, the merger occurs within 20\,Myr
of the initial star formation time (for main-sequence star mergers,
the distribution can extend to 100\,Myr).  These objects will likely 
trace star formation.

\subsection{Mass and Orbital Separation Distributions}

Our populations demonstrate that compact remnant mergers can occur for
a wide range of star characteristics: mass, evolutionary phase and,
hence, orbital separation at the time of the merger.  In
section~\ref{sec:observations}, we will use these characteristics to 
determine observable features of these mergers.  Before we do so, 
let's review the characteristics of these populations.

\begin{figure}[htbp]
\epsscale{0.70}
\plotone{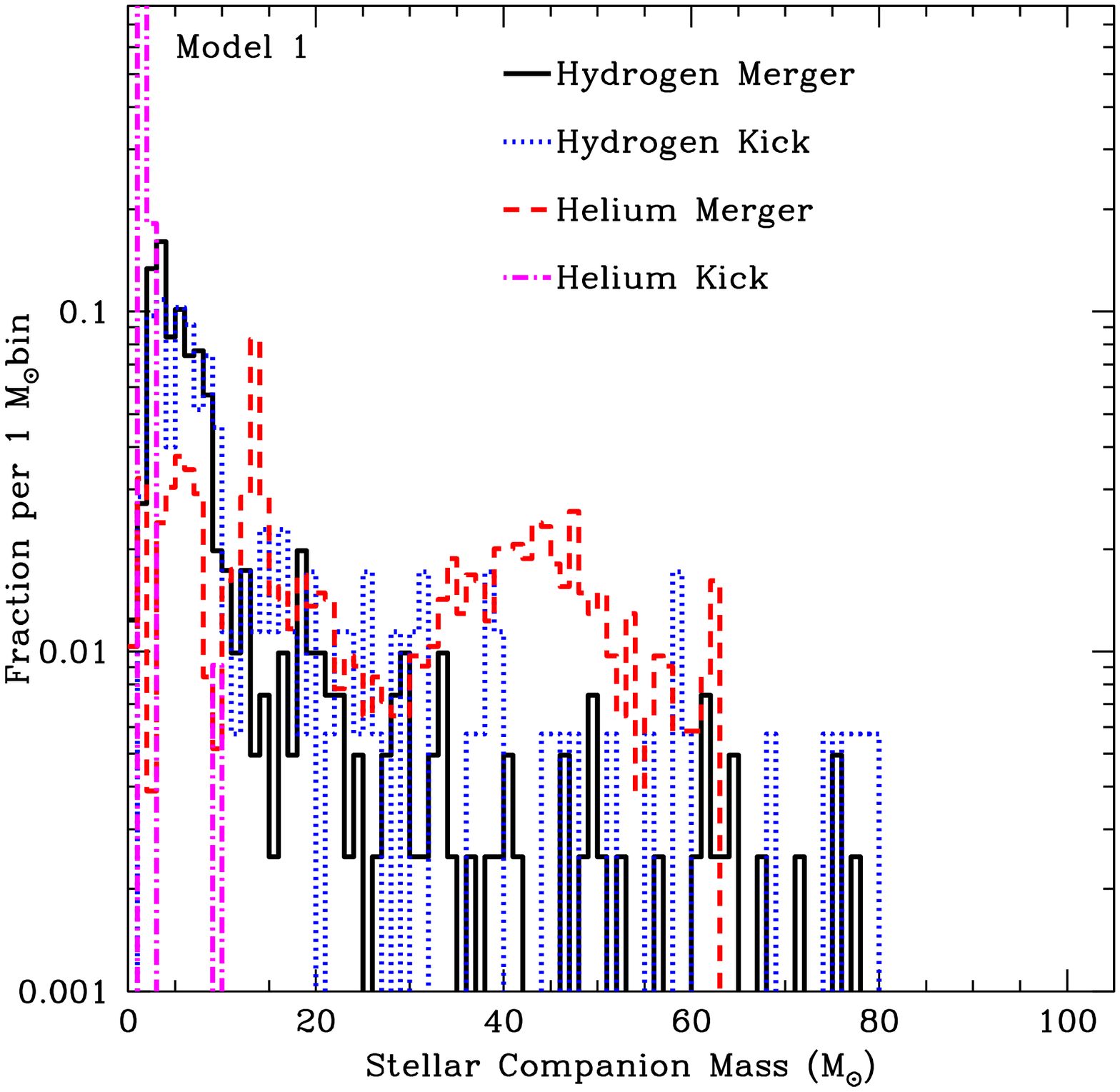}
\plotone{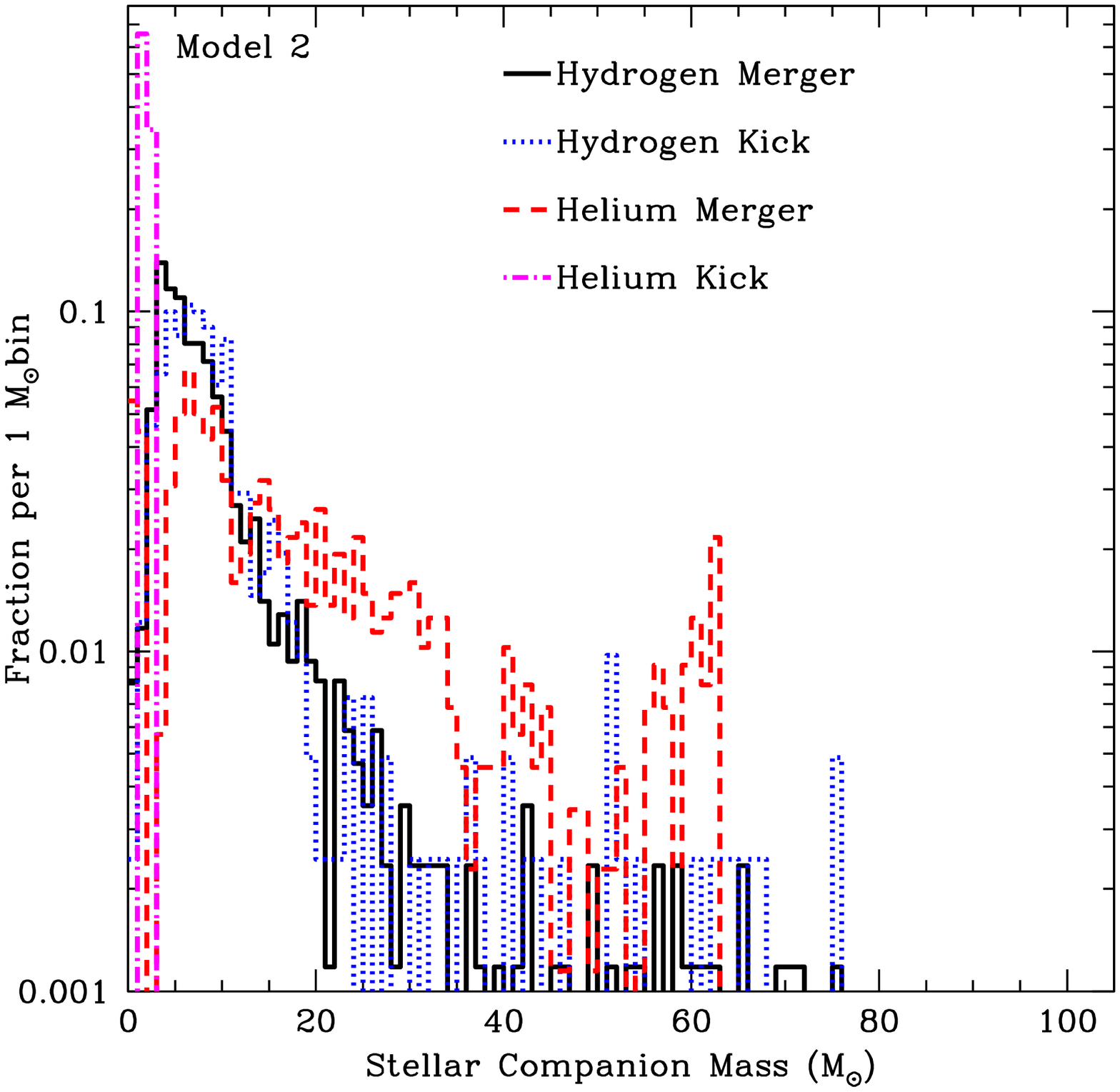}
\plotone{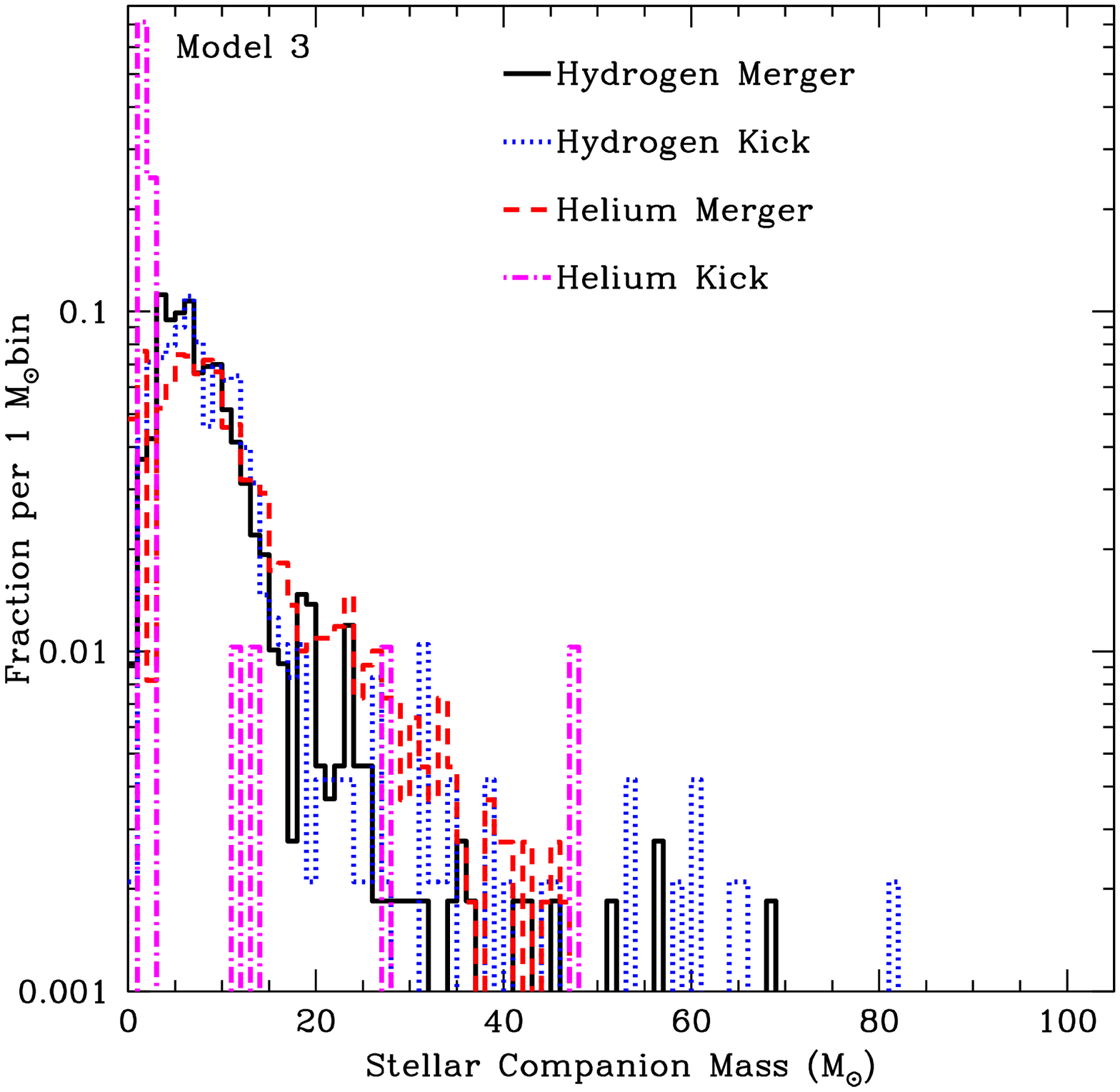}
\plotone{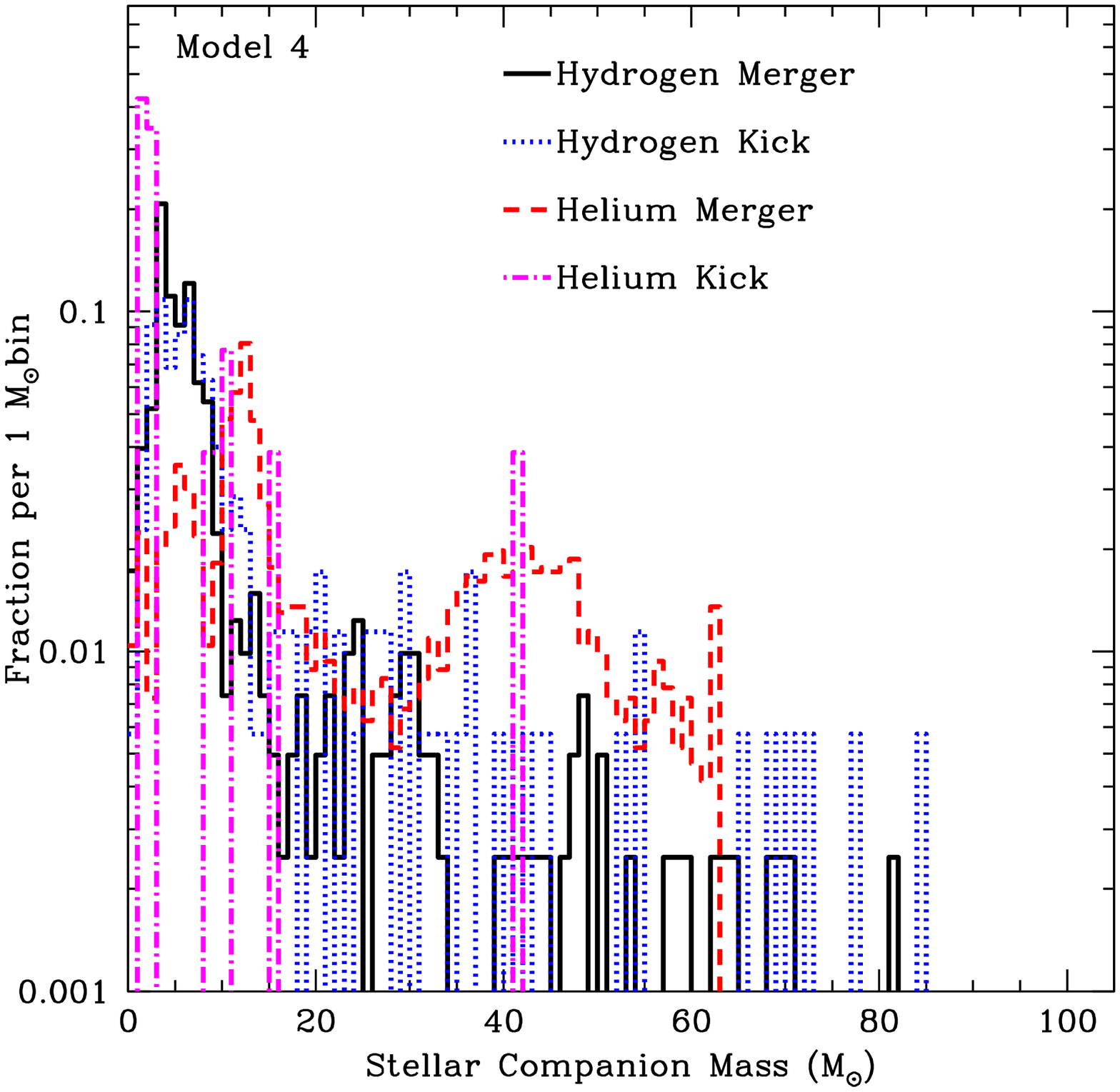}
  \caption{Companion star mass distribution for 4 different
    populations: those with hydrogen cores merging through mass
    transfer, those with hydrogen cores merging when the compact
    remnant is kicked into the companion, those with helium cores
    merging through mass transfer, and those with helium cores merging
    when the compact remnant is kicked into the companion.  Except for
    helium merger systems (which have companion masses ranging up to
    60\,M$_\odot$, often with a secondary peak at these high masses),
    most of the systems have companion masses peaking strongly at low
    masses.}
  \label{fig:mass}
\epsscale{1.0}
\end{figure}

\begin{figure}[htbp]
\epsscale{0.70}
\plotone{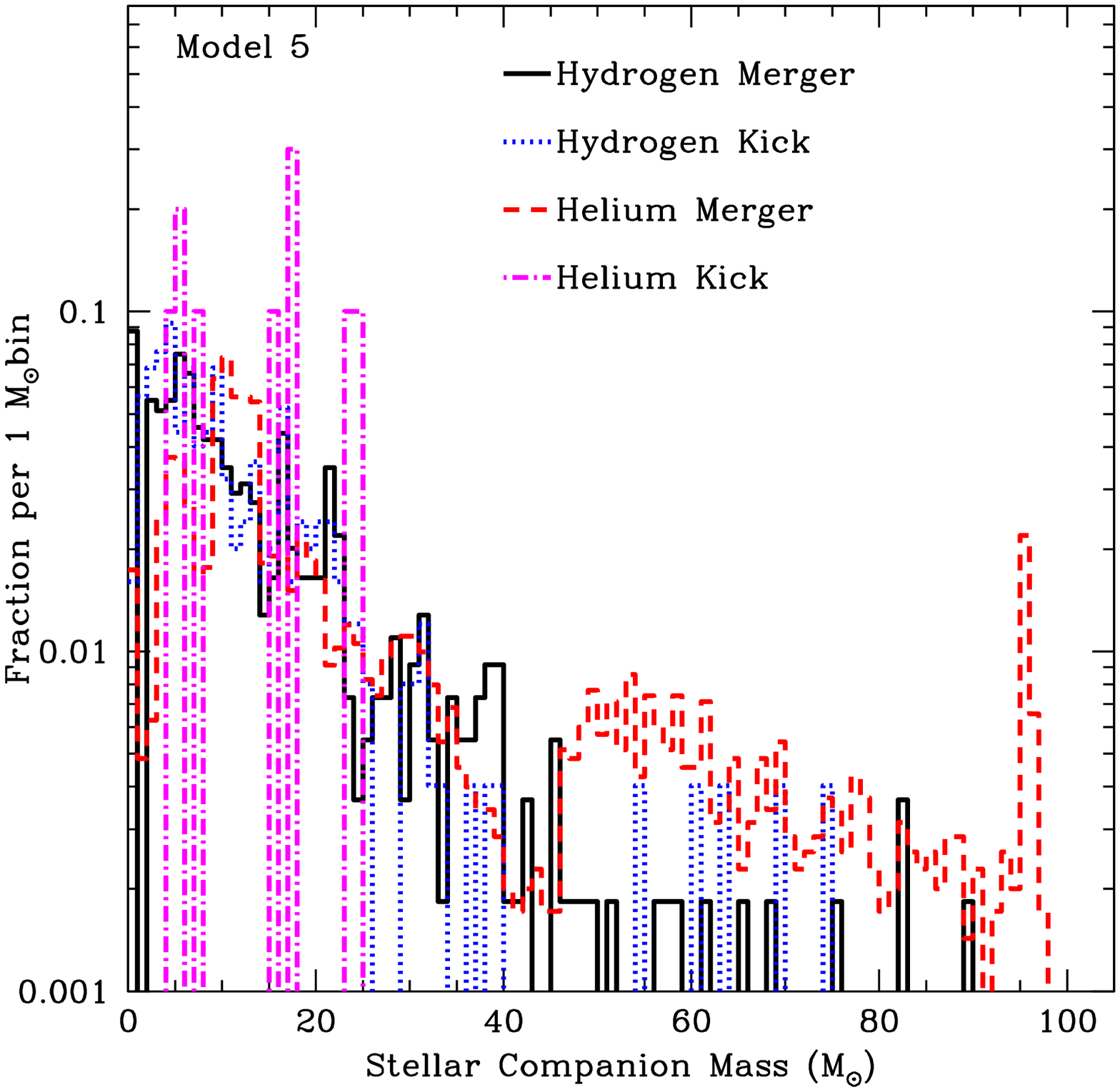}
\plotone{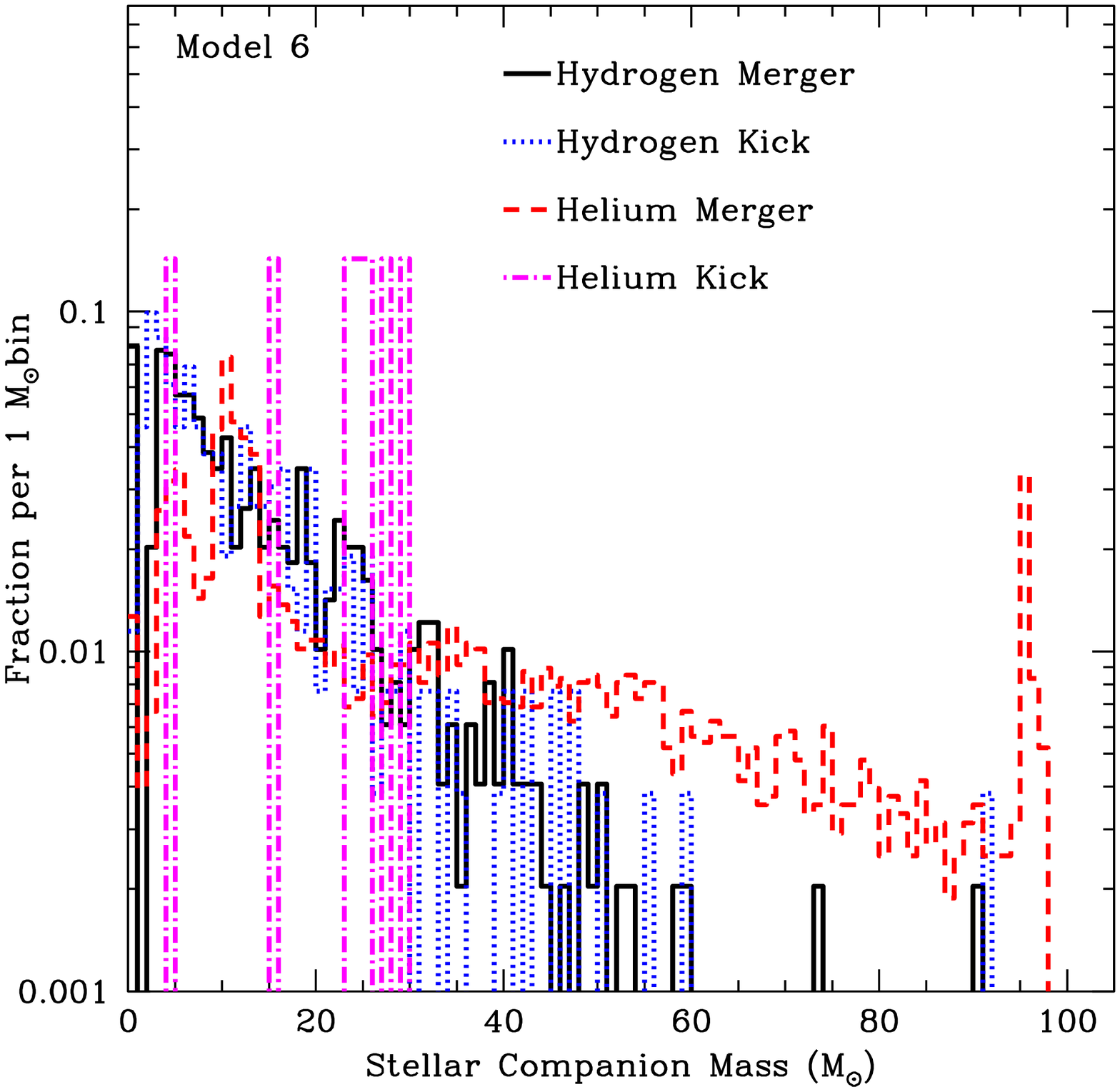}
  \caption{Companion star mass distribution for 4 different
    populations (see fig.~\ref{fig:mass}).  Note that for low
    metallicity systems, the helium merger population has a 
    much broader high-mass tail, producing mergers with helium 
    cores in excess of 90\,M$\odot$.}
  \label{fig:masslowz}
\epsscale{1.0}
\end{figure}

Figures~\ref{fig:mass},\ref{fig:masslowz} show the distribution of
companion masses for 4 different merger populations: those with
hydrogen cores merging through mass transfer, those with hydrogen
cores merging when the compact remnant is kicked into the companion,
those with helium cores merging through mass transfer, and those with
helium cores merging when the compact remnant is kicked into the
companion.  As we might expect with our steep initial mass function,
the companion masses are peaked to lower mass (below 20\,M$_\odot$).
However, the distribution is broad, extending beyond 60\,M$_\odot$.
He-merging systems have a secondary peak at higher masses and the most
massive helium cores will come from this peak.  For low-metallicity
stars, this secondary peak occurs nearly at 90\,M$_\odot$.

\begin{figure}[htbp]
\epsscale{0.70}
\plotone{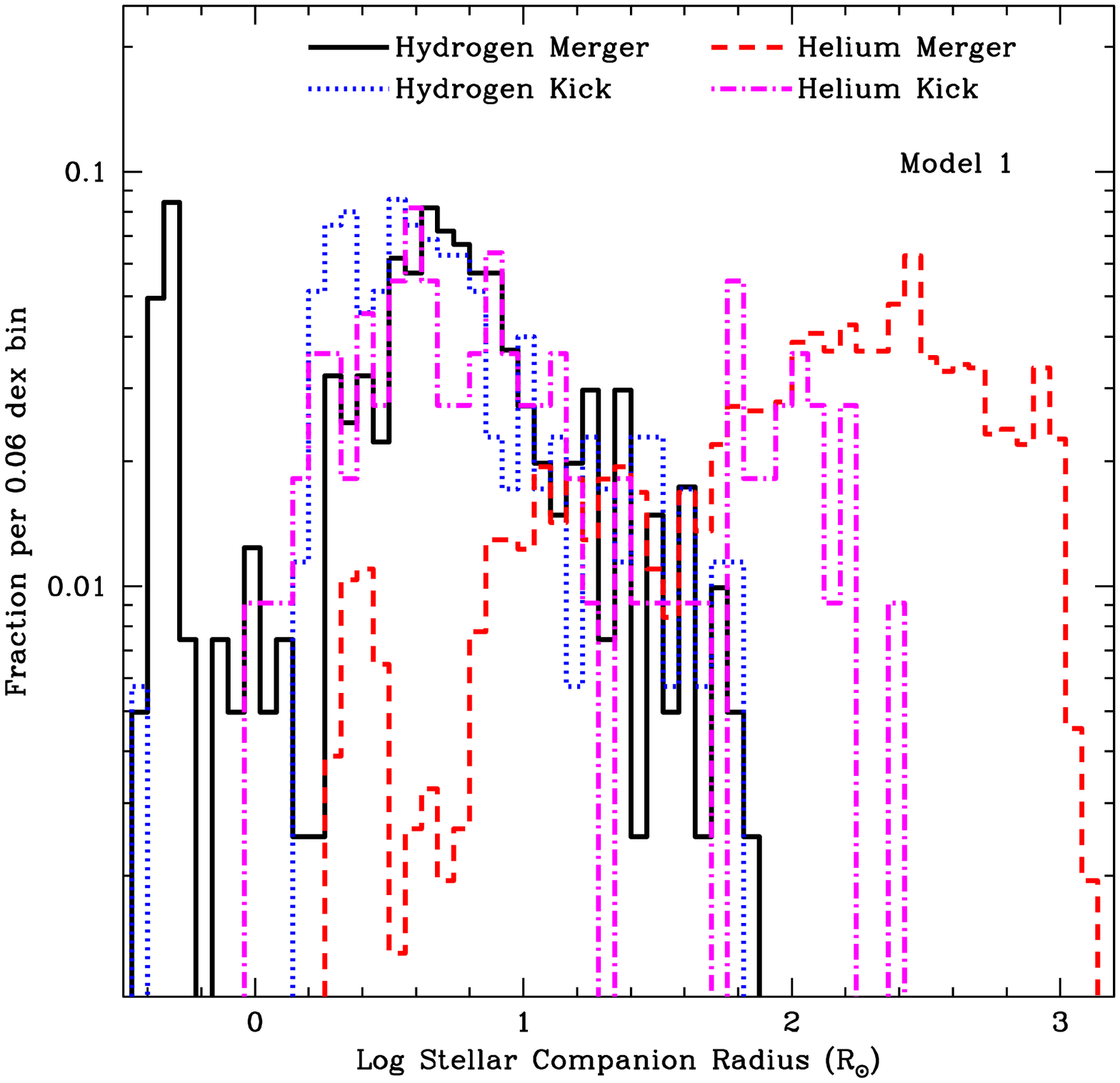}
\plotone{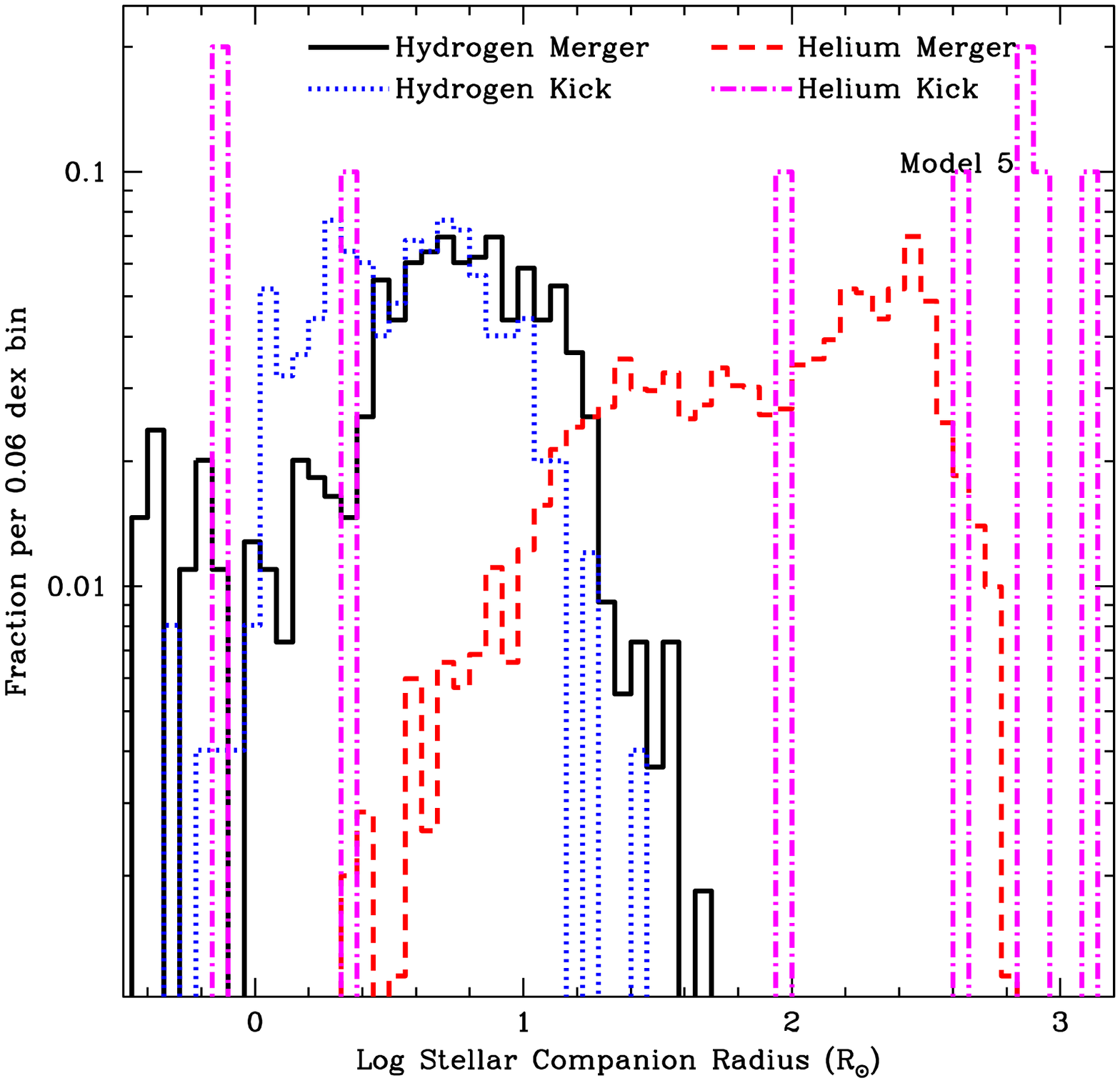}
\plotone{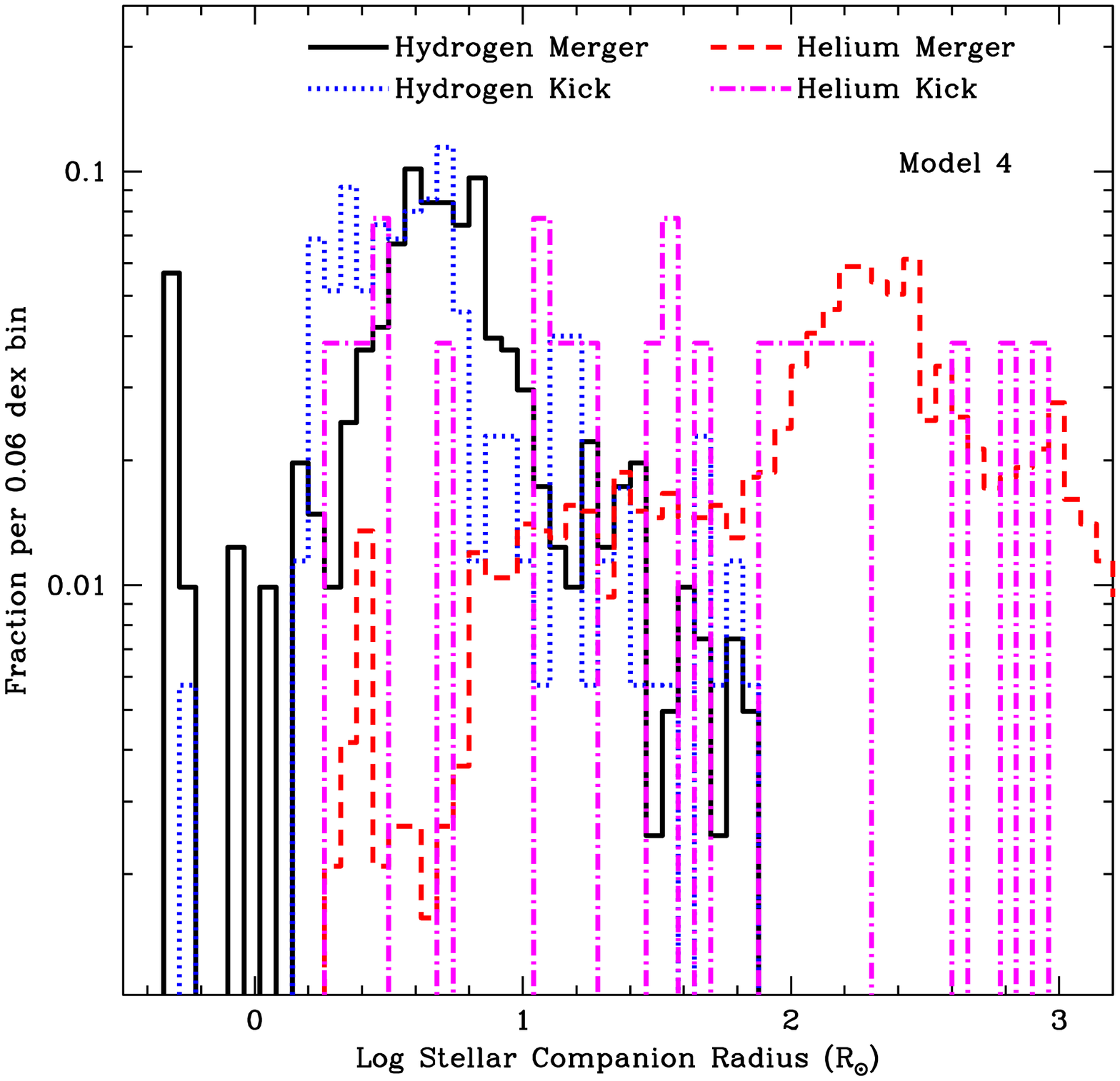}
\plotone{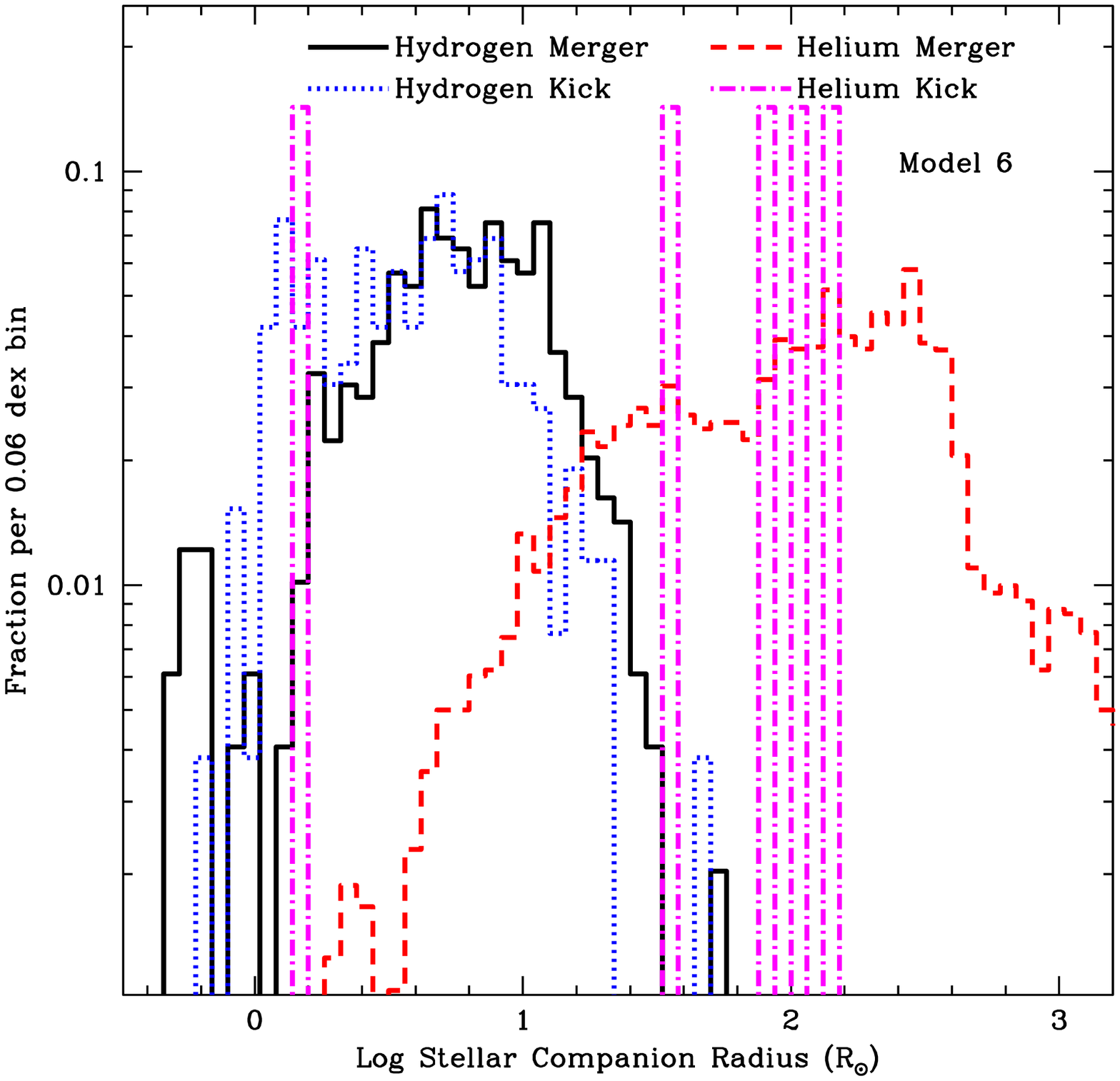}
  \caption{Companion star radius distribution for 4 different
    populations in Figure~\ref{fig:mass}.  The evolved stars (with
    helium cores) have much higher radii (up above 5 A.U.).  Because 
there is very little variation between models, we show only 4 of our 
models.}
  \label{fig:rad}
\epsscale{1.0}
\end{figure}

Similarly, the distribution of companion radii is extremely broad
(Fig.~\ref{fig:rad}).  As expected, the evolved stars (with helium
cores) can be much larger (up beyond 5 A.U.).  The helium core sizes 
tend to lie in a narrow range between 0.5-3\,R$_\odot$.

The wide variation in these systems will produce a wide variation 
in the observational features of these explosions.

\section{Observational Comparisons}
\label{sec:observations}

The characteristics of the compact binaries (companion masses,
evolutionary phases, radii and orbital separations) allow us to
estimate both the explosion luminosities and spectral features.  

\subsection{Luminosities}

For explosion luminosities, we will divide the discussion into two
categories based on the accretion rate.  If the system has a massive
helium core ($> 4M_\odot$), the accretion is sufficient to power a
strong black-hole accretion disk engine discussed in classical GRBs.
In these cases, even the neutrin-annihilation powered engine may be
strong.  These high accretion-rate systems fit the classic Helium
Merger model proposed by Fryer \& Woosley (1998).  We then study
low-mass helium core and hydrogen core systems.

\subsubsection{Massive Helium-core Merger:  GRB-like}
\label{sec:hemerger}

Our population synthesis models predict a wide set of merging systems
involving a compact remnant (neutron star or black hole) with its
companion.  As we showed in section~\ref{sec:poprate}, 8-50% of these
are mergers with evolved stars with helium core masses above
4\,M$_\odot$.  With a rate of 10-120\,Myr$^{-1}$, mergers with massive
helium cores can make up a significant fraction of our GRB population.
In this section, we use our current understanding of disk-engine models 
to predict explosion luminosities from the helium core mass distribution.

\begin{figure}[htbp]
\plotone{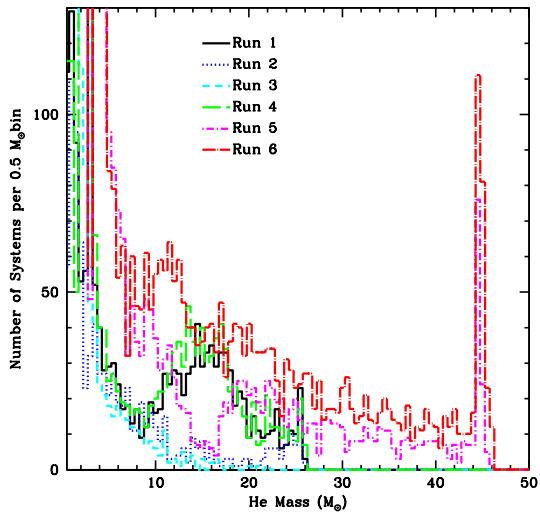}
  \caption{Mass distribution of helium cores in our population of
    he-mergers.  The two largest peaks exist for low-mass 
helium cores: 1.4 and 3.4\,$M_\odot$.  But the distribution is 
widespread (half of all merging systems involve helium cores 
with masses above 4\,$M_\odot$).}
  \label{fig:hemassdist}
\end{figure}

Zhang \& Fryer (2001) studied the accretion onto a neutron star during
an inspiral.  They found that, the more massive the helium core, the
higher the accretion rate (and, ultimately, the more energetic GRB
explosion).  Our population synthesis models provide us with the
distributions of helium core masses at the onset of the merger is
shown in Figure~\ref{fig:hemassdist}.  Although the helium core masses
peak at low masses, the distribution of helium core masses for solar
metallicy models extends beyond 25\,$M_\odot$ and roughly half of the
systems have helium cores with masses above 4\,$M_\odot$.  At low
metallicities, the maximum helium core mass extends from 25\,M$_\odot$
up to 45\,M$_\odot$.

We can estimate the accretion rate onto the compact remnant ($\dot{M}_{\rm rem}$):
\begin{eqnarray}
\dot{M}_{\rm rem} & = & \eta M_{\rm He}/t_{\rm free-fall} = \nonumber\\ 
&& \eta [2 M_{\rm He} \sqrt{2.0 G (M_{\rm He} + M_{\rm rem})}]/[\pi R_{\rm He}]
\end{eqnarray}
where $t_{\rm free-fall}$ is the free-fall timescale, $M_{\rm He}$,
$R_{\rm He}$ are the helium core mass and radius respectively, $M_{\rm
  rem}$ is the compact remnant mass and $G$ is the gravitational
constant.  The accretion rate is less than the typical free-fall time
because the material has thermal pressure and considerable angular
momentum.  To account for this, we include an efficiency parameter
$\eta$.  Fitting to the Zhang \& Fryer (2001) results, we find the
value of $\eta$ to be roughly 15\%.

\begin{figure}[htbp]
\plotone{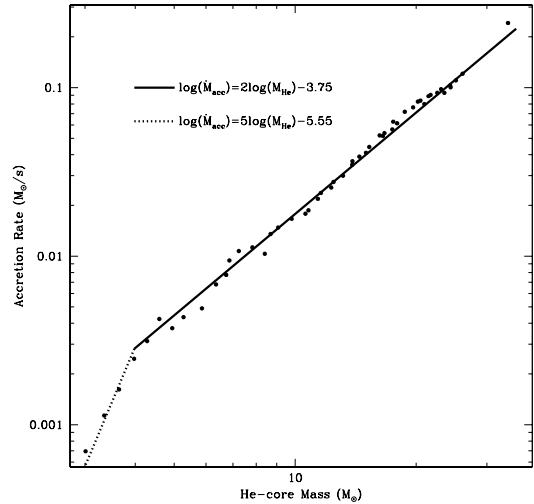}
  \caption{Average accretion rate as a function of helium core 
mass.  The dots show results using the helium cores from the 
Woosley et al. (2002) models.  The accretion rate data is fit by a 
2-piece power law of the helium core mass.  Below 4\,M$_\odot$, 
the power increases dramatically, driving the accretion rate 
down significantly as the helium core mass decreases.}
  \label{fig:mdot}
\end{figure}

By using the helium core radii as a function of mass from the Woosley
et al. (2002) pre-collapse helium cores, we derive an accretion rate as a
function of helium core mass is shown in figure~\ref{fig:mdot}.  These 
helium cores are evolved (at collapse), but their radii are not too 
different from the unevolved cores used by Zhang \& Fryer (2001).  This 
mass loss rate is fairly well fit by a 2-part power law:
\begin{eqnarray}
\log(\dot{M}_{\rm rem}) &=& 2 \log(M_{\rm He}) - 3.75 \; {\rm if} \, M_{\rm He}>4.0 \nonumber\\
  && 5 \log(M_{\rm He}) - 5.55 \; {\rm otherwise}.
\end{eqnarray}
For helium cores with masses below 4\,$M_\odot$, the accretion rate drops 
rapidly.  These low-mass cores are less likely to drive strong explosions, 
and for the purposes of our discussion of helium mergers, we will focus 
on those helium cores with masses above 4\,$M_\odot$.

\begin{figure}[htbp]
\plotone{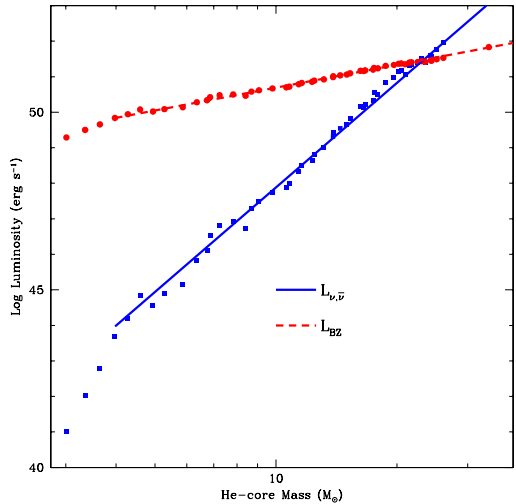}
  \caption{The luminosity versus helium core mass for our 
two power sources:  neutrino annihilation ($L_{\nu,\bar{\nu}}$), 
Blandford-Znajek emission ($L_{\rm BZ}$).  The stronger dependence 
of neutrino annihilation on the density and temperature (and 
hence accretion rate) means that the luminosity from neutrino 
annihilation is extremely sensitive to the helium core mass.}
  \label{fig:lum}
\end{figure}

Zhang \& Fryer (2001) estimated the explosion luminosity as a function of
accretion rate for a helium merger GRB assuming both the
Blandford-Znajek~\citep{Bla77} and neutrino annihilation
mechanisms.  For neutrino annihilation, we use the formula from Zhang 
\& Fryer(2001):
\begin{eqnarray}
\log[L_\nu,\bar{\nu} ({\rm erg s^{-1}})]  & \approx &  43.6 + 4.89
\log[\dot{M}_{\rm rem}/(0.01 M_\odot s^{-1})] \nonumber\\
 &&  + 3.4a_{\rm rem}
\end{eqnarray}
where $a_{\rm rem} \equiv (J_{\rm rem} c)/(G M_{\rm rem}^2)$, $J_{\rm
  rem}$ is the angular momentum of the remnant and $c$ is the speed of
light.  Values of $a_{\rm rem}$ for these mergers tend to lie in the
range of 0.75-0.95.  We will assume 0.9 for our estimates.  This
estimate assumes a roughly 3\,M$_\odot$ black hole which, for the
purposes of this study, is appropriate.  However, other formulae exist
that will produce different energy distributions, e.g. Zalamea \&
Beloborodov (2011).  Our prescription produces a broader energy
distribution than Zalamea \& Beloborodov (2011) and, to demonstrate
the range of energy distributions, we will focus on these formulae.
However, one should bear in mind that the distribution of neutrino
annihilation outbursts may be more narrow than predicted by our
results.

The comparable equation for Blandford-Znajek 
emission is~\citep{Popham99}:
\begin{equation}
\log[L_{\rm BZ} ({\rm erg s^{-1}})] \approx 50.0 + 2 \log[a_{\rm rem} 
M_{\rm rem} B_{\rm mag}]
\end{equation}
where $B_{\rm mag}$ is the magnetic field strength.  As with Zhang \& Fryer (2001), 
we will assume the magnetic energy density is equal to the kinetic energy density:
\begin{equation}
B^2 = 4 \pi \rho v^2 \approx \dot{M}_{\rm rem}c/r_g^2
\end{equation}
where we have approximated the density ($\rho$) and velocity ($v$) of
the material using the values at the event horizon: $r_g = 2 G M_{\rm
  rem}/c^2$.  Again, we set $a_{\rm rem}$ to 0.9 for our estimates.  
The corresponding luminosity for both these models as a function  
of helium core mass is shown in figure~\ref{fig:lum}.  Neutrino annihilation 
is much more sensitive to the helium core mass and leads to a broader spread 
of explosion energies.

\begin{figure}[htbp]
\plotone{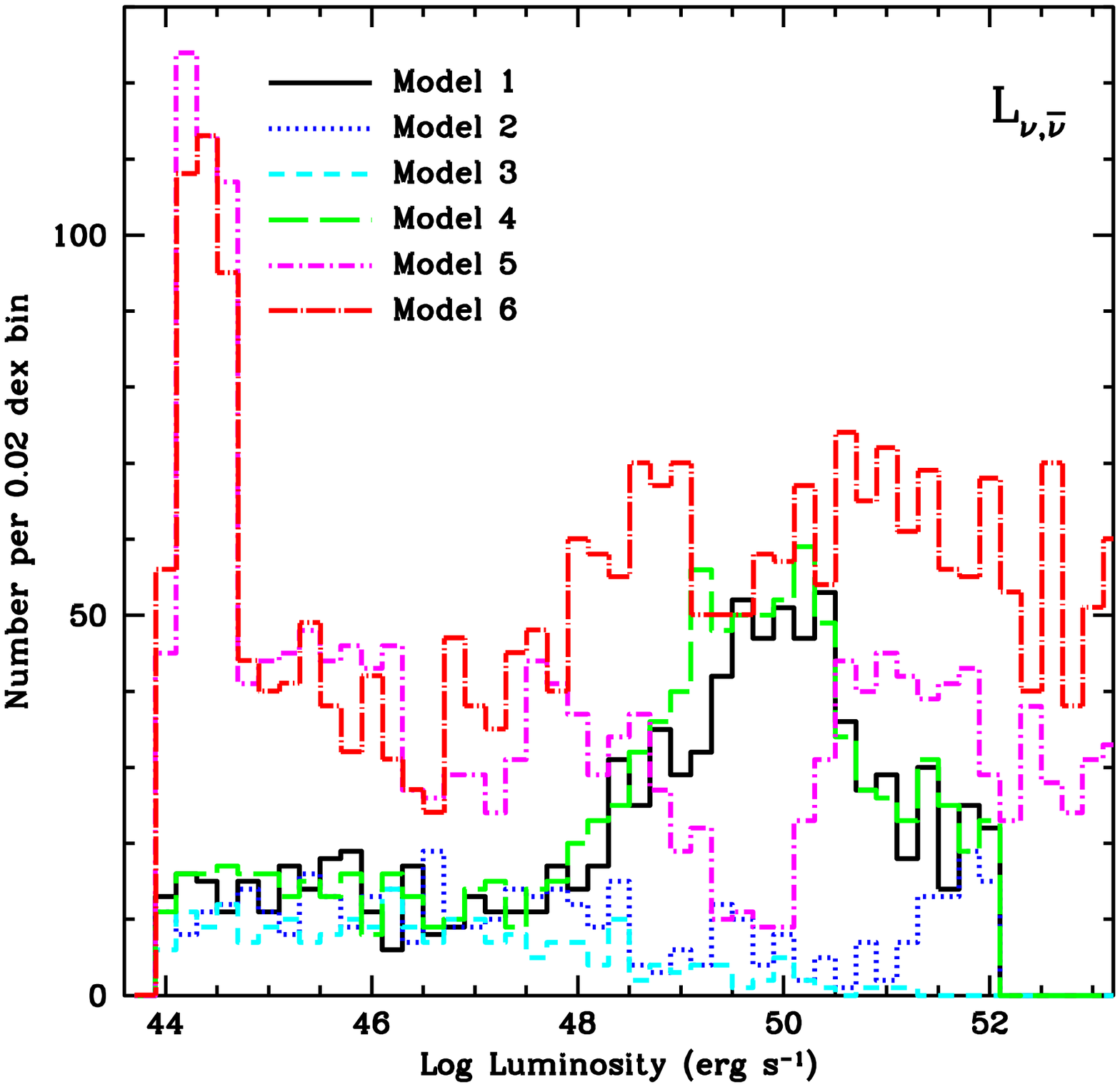}
\plotone{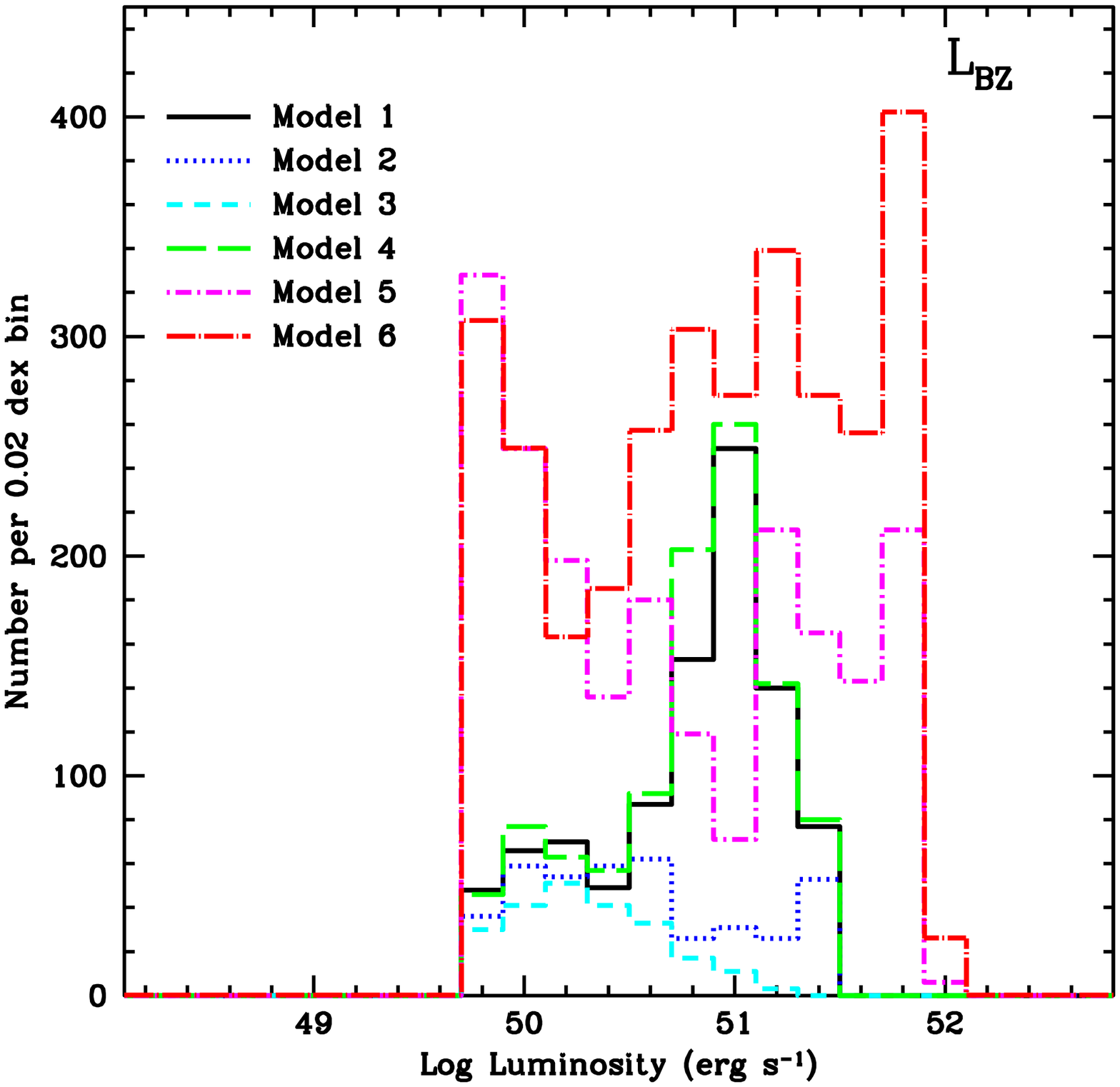}
  \caption{The distribution of luminosities from the helium mergers
    from our population synthesis calculations.  The neutrino
    annihilation model's strong sensitivity on the helium core mass
    leads to a broad distribution in the explosion energies.}
  \label{fig:lumdist}
\end{figure}

These formulae allow us to estimate the luminosity of the outburst
from our helium merger population.  The distributions for both the
neutrino annihilation and Blandford-Znajek models are shown in
figure~\ref{fig:lumdist}.  As expected, the sensitivity of the
neutrino annihilation emission model leads to a broad spectrum of
explosion luminosities.  Our simplified Blandford Znajek model
predicts a relatively narrow range of luminosities from
$10^{49}-10^{51}$\,ergs\,s$^{-1}$.

\begin{figure}[htbp]
\plotone{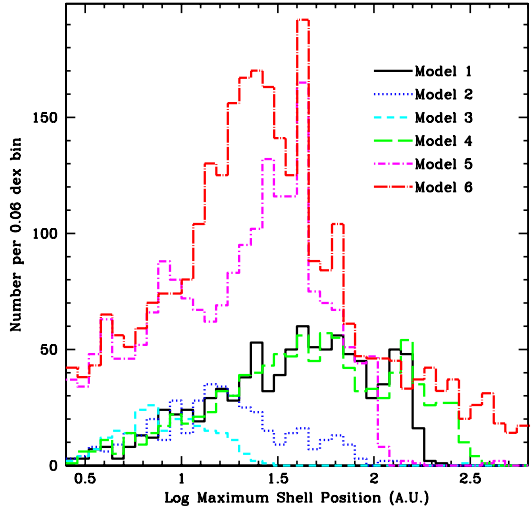}
  \caption{The distribution of outer positions of the helium 
merger ejecta from our population synthesis models.}
  \label{fig:rshelldist}
\end{figure}

The merger process ejects the hydrogen atmosphere of the star in a
shell of material with an density enhancement along the orbital plane.
It was this shell that Th\"one et al. (2011) exploited to explain the
peculiar features of the Christmas burst.  They argued that
interactions of the jet with this shell produced the late-time, large
radius blackbody emission.  We estimate the outer extent of this shell
using the same analysis of Th\"one et al. (2011): assume the inspiral
occurs in $\sim$3 orbits and the material ejected in this common
envelope phase moves outward at the escape velocity.  Although these
two assumptions are not strictly true, we can not yet model this
process exactly (e.g. Passey et al. 2011, Ivanova et al. 2012).  These
assumptions approximate the ultimate result within a factor of 2 in
most cases.  Using these assumptions, the corresponding distribution
of shell positions for our populations are shown in
figure~\ref{fig:rshelldist}.

\begin{figure}[htbp]
\epsscale{0.7}
\plotone{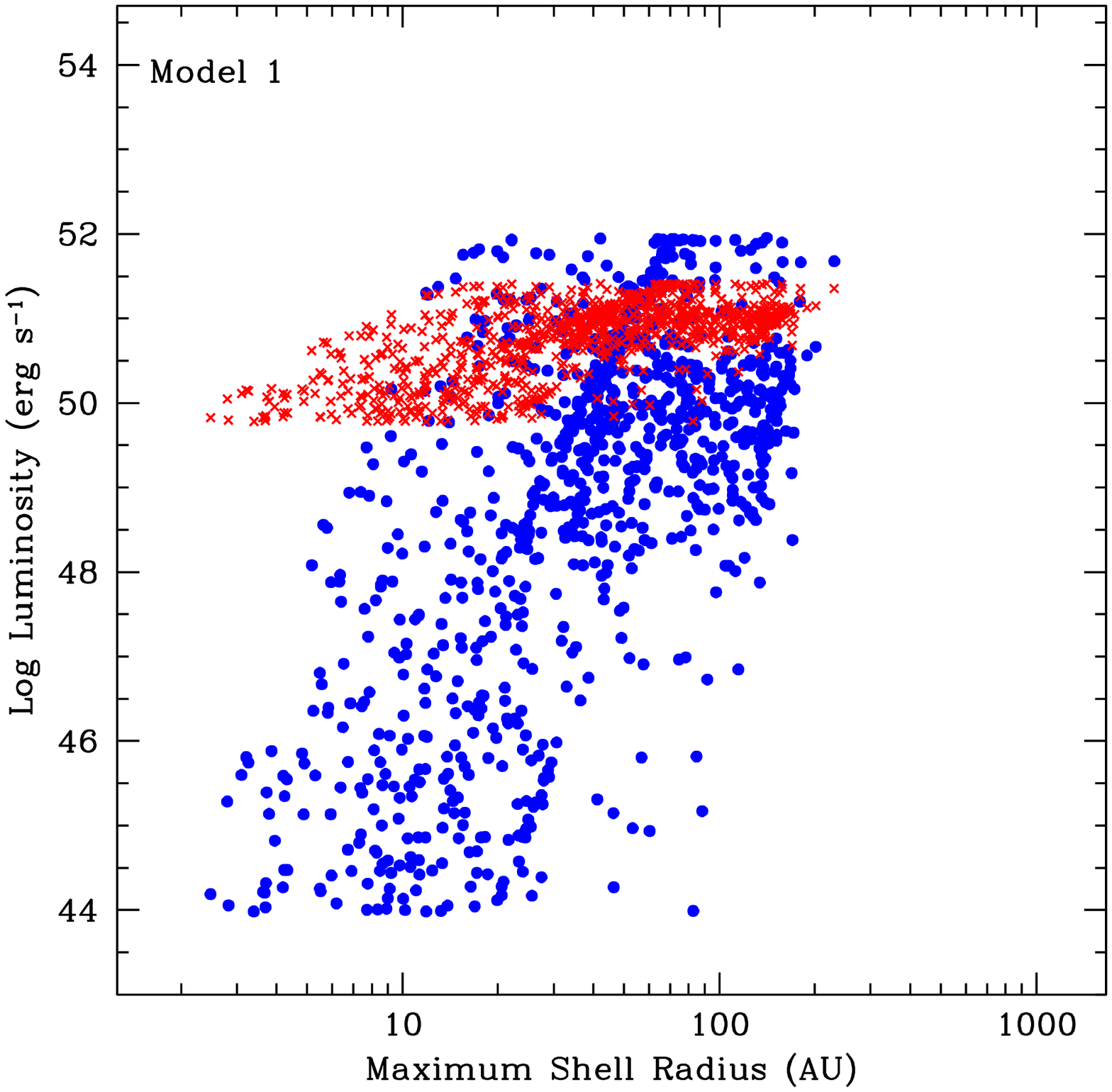}
\plotone{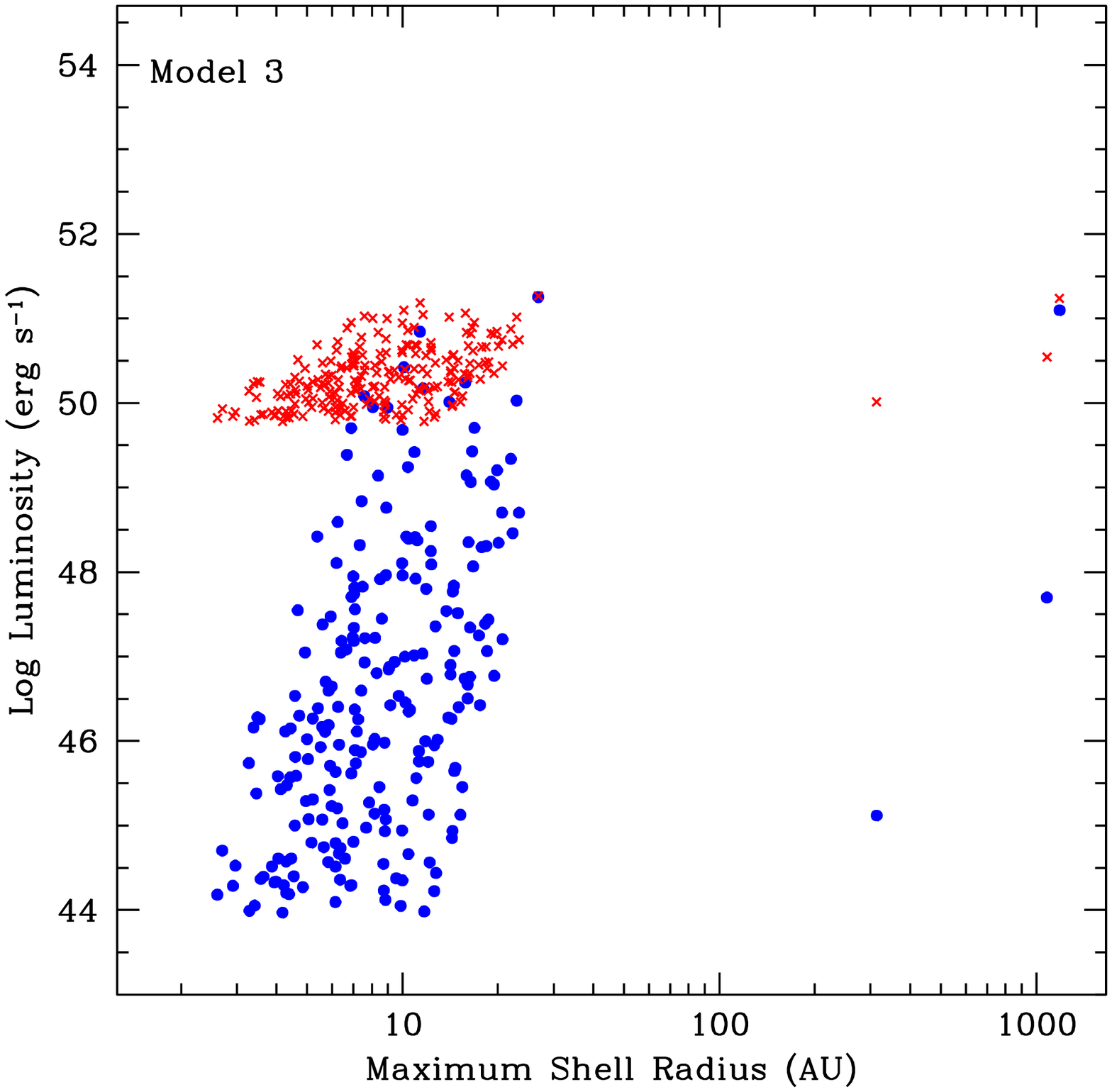}
\plotone{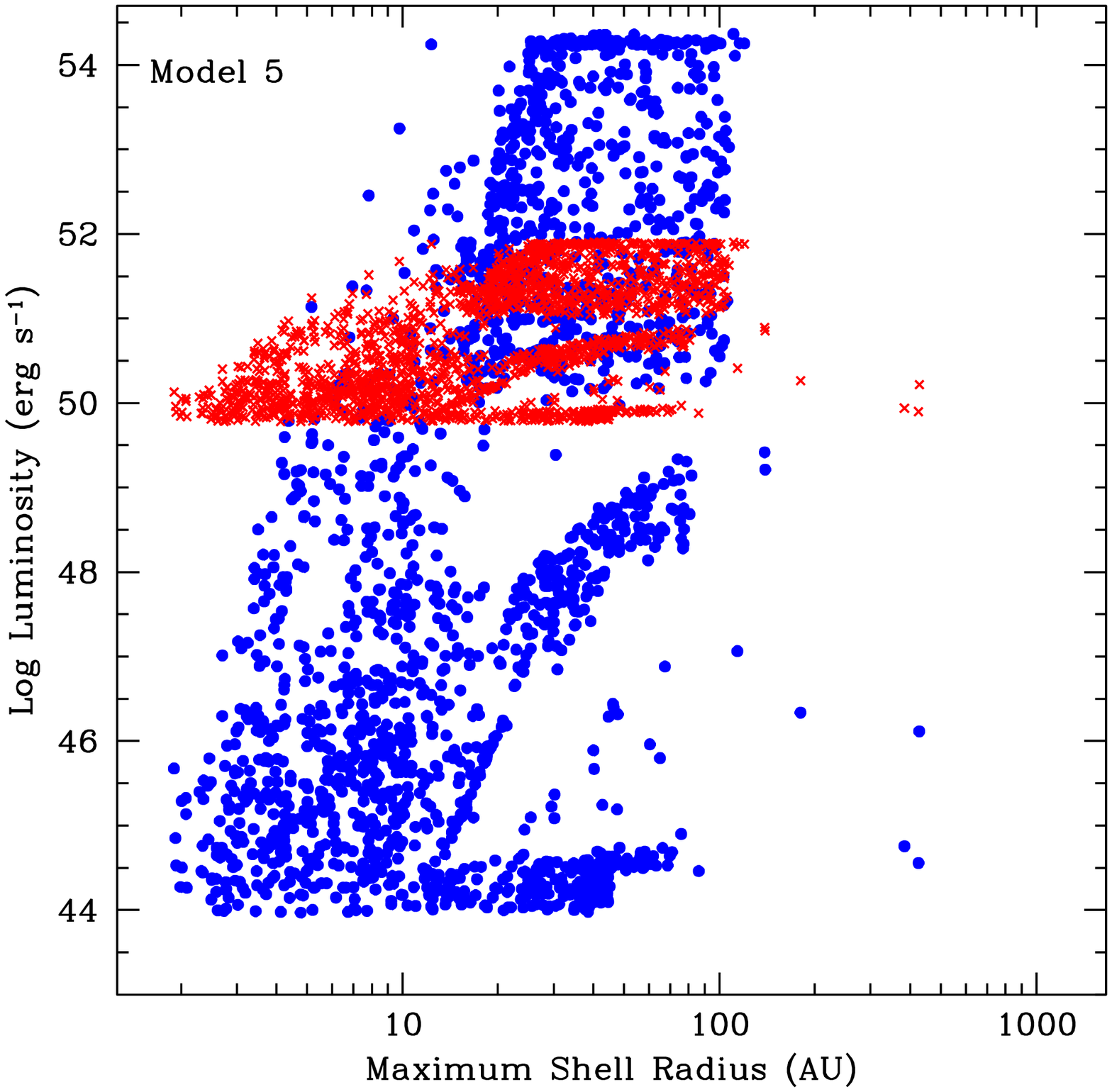}
\plotone{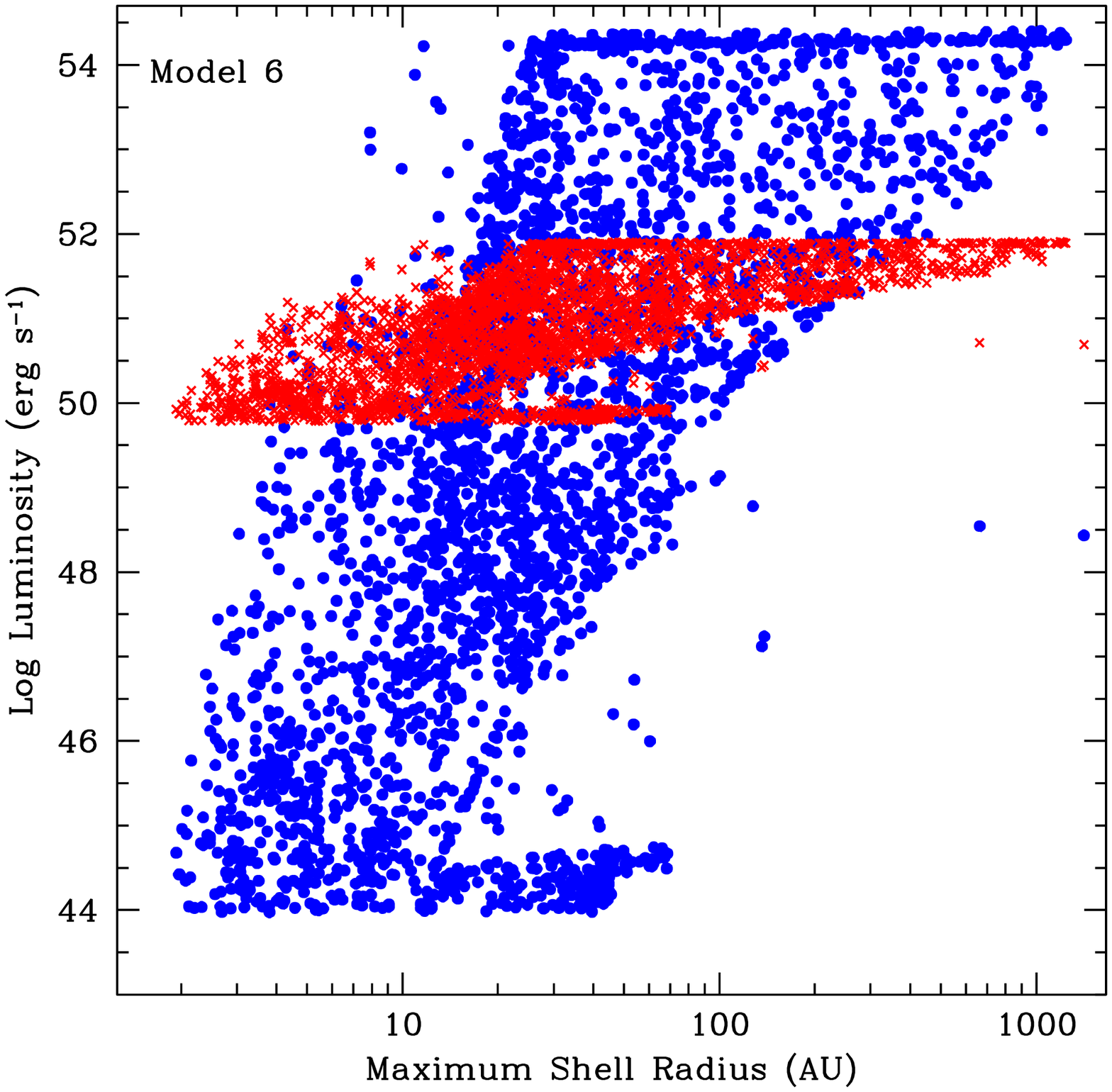}
  \caption{Luminosities versus outer shell radii for our 
helium merger population.  Our Blandford-Znajek, magnetic 
field, engine (crosses) tends to produce stronger explosions in 
general, but the strongest explosion energies arise from 
our neutrino annihilation model (circles).  The strongest energies tend 
to have the widest separations, but a there is considerable 
scatter in the maximum radius/luminosity relation.}
  \label{fig:lumvshell}
\epsscale{1.0}
\end{figure}

Our population helium mergers predicts a range of luminosities versus
maximum shell radii (Fig.~\ref{fig:lumvshell}).  There is considerable 
scatter in the relationship between the outer shell radii and luminosity.  
But there is a trend that the most luminous explosions have the furthest 
shell distributions. 

\subsubsection{Supernova-like Explosions:  Low-Mass Helium Cores and Main Sequence Mergers}

If the compact object merges with a lower-mass helium core or a
main-sequence star, the accretion rate onto the remnant will be much
lower.  For these systems, the black hole accretion disk engine 
is unlikely to be valid and it is more difficult to produce even 
a rough estimate of the explosion energy.  

Before the common envelope phase, many of our systems are likely 
to be X-ray binaries.  At the onset of the common envelope phase, 
the X-ray binary can evolve, becoming more powerful.  The system 
may even begin to produce strong jets.  These so-called, micro-quasars, 
have been observed in numerous systems (see Mirabel 2007 for a review).

What happens as the compact remnant spirals into its companion is 
more difficult to determine.  In these mergers, the accretion rate 
is slow enough that the neutron star may not initially collapse to 
form a black hole.  The accretion rate remains high enough that 
any neutron star magnetic field is likely to be buried.
Even without magnetic fields, these accretion events can produce
explosions on par with normal supernova~\citep{Fry06a,Fry09}.  

Presumably, once the compact remnant collapses to a black hole, 
the accretion disk will produce a jet.  In these hydrogen or 
low-mass helium cores, the accretion rate is insufficient for 
the neutrino annihilation model to work, but we still do not 
understand the black hole accretion disk mechanism well enough 
to predict whether such systems will produce jets with explosions 
at GRB levels or at sub-supernova levels.

\begin{figure}[htbp]
\plotone{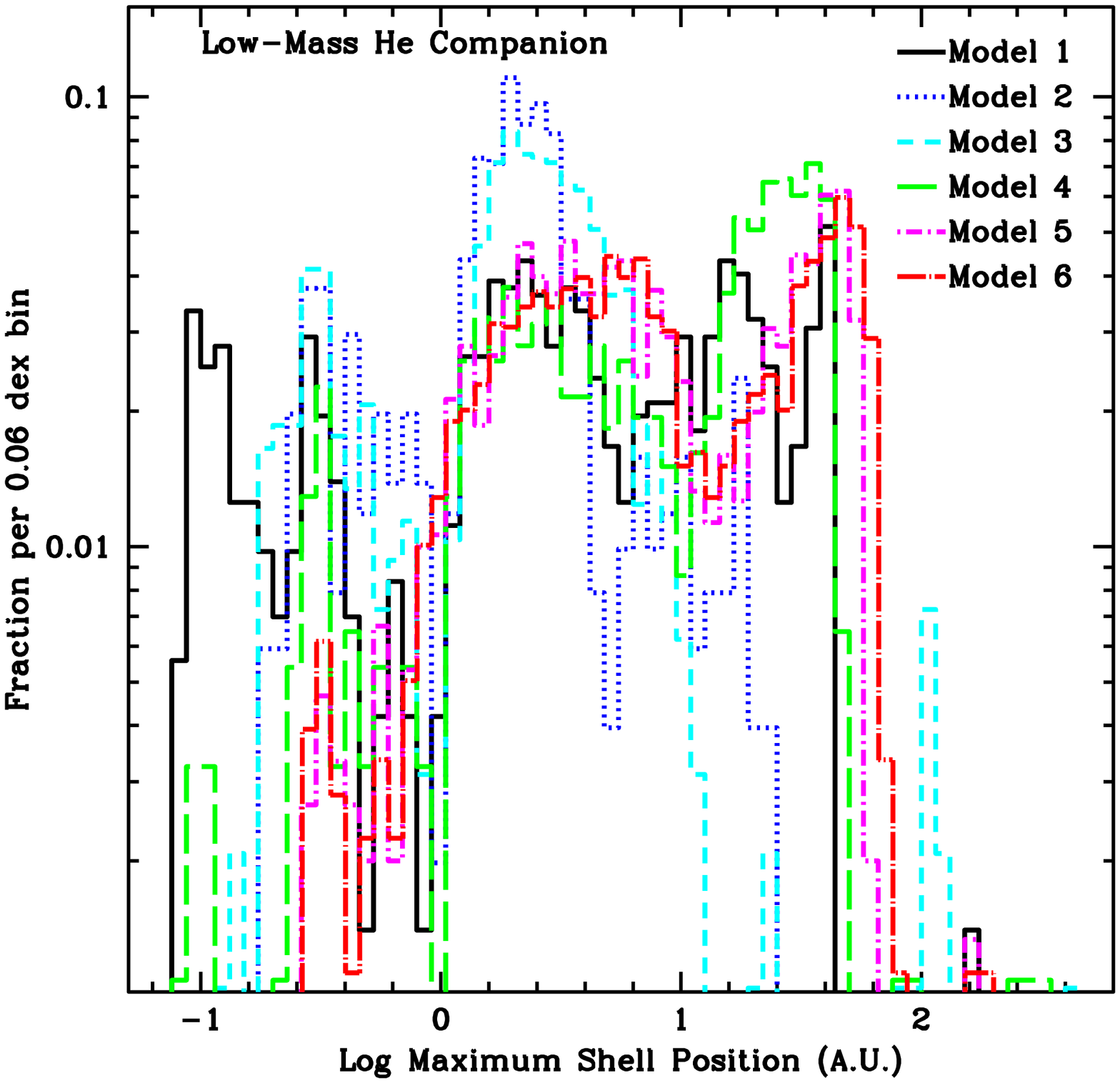}
\plotone{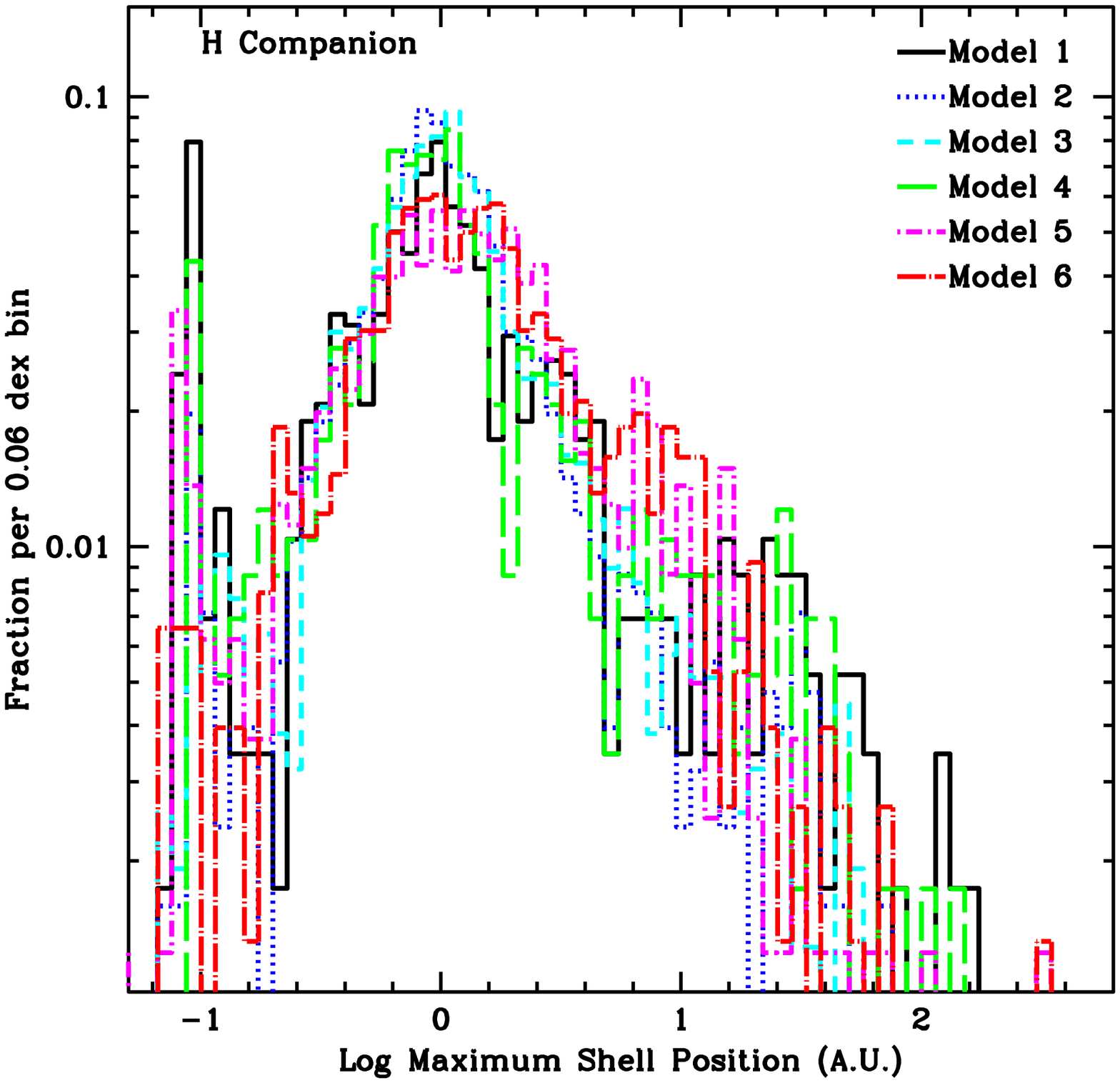}
  \caption{The distribution of outer positions of our low-mass 
helium cores and hydrogen stars.  The distribution is much 
broader than the high-mass helium cores, with many more systems 
with quite low outer shell radii.  The broad range of shell 
separations from these systems is in large part due to the fact 
that a sizable fraction of these systems (30\% of hydrogen stars, 
17\% of low-mass helium stars) arise from systems where the 
compact remnant is kicked into its companion.}.
  \label{fig:rhydist}
\end{figure}

\subsection{Nucleosynthetic Yields}

Two sites have been identified in long-duration GRBs to produce
$^{56}$Ni: explosive nucleosynthesis in the shock (this shock burning
is what produces the $^{56}$Ni in supernovae), or disk-wind
nucleosynthesis~\citep{Sur06, Sur11}.  

Explosive nucleosynthesis produces $^{56}$Ni in the silicon and oxygen
layers of an evolved core in normal supernovae or classic collapsars.
Without an evolved core, it is difficult for the helium merger model to
produce significant $^{56}$Ni from shock interactions.  Unless
considerable material is processed and ejected in the disk, very
little $^{56}$Ni is produced.  Like the fallback GRBs discussed in
\cite{Fry07b}, these GRBs could produce low-nickel yield outbursts.
Although this may not affect the peak of the light-curve (in most of
these systems, shock heating dominates the light-curve at peak: Fryer
et al. 2007b, Fryer et al. 2009), it does mean that if explosive
nucleosynthesis is the dominant source of $^{56}$Ni, these explosions
will not have typical $^{56}$Ni decay tails at late times.

However, if disk winds dominate the $^{56}$Ni production, the
nucleosynthetic yields of GRB-like he-mergers will be
indistinguishable from collapsar long-duration GRBs.

\subsection{Spectra}

Contamination from the ejected hydrogen envelope is the strongest
constraint for he-mergers.  However, the hydrogen distribution in the
helium merger model than in typical stellar envelopes.  Even though
the hydrogen remains within $10^{15}$\,cm of the central engine, the
distribution is very asymmetric.  \cite{Pas12} found that 90\% of the
mass ejected was distributed within 30$^{\circ}$ of the equator.  It
is more difficult to accurately describe the distribution of the
remaining 10\%.  Combined with the fact that the explosion is also 
asymmetric, it is possible to sweep up very little mass.

How low must the swept-up hydrogen mass be?  For symmetric explosions,
\cite{Hac12} found that roughly 0.1 solar masses of hydrogen could be
hidden in the explosion.  For asymmetric supernovae, this number may
be a bit greater.  This would require only 1-2\% of the
stellar-envelope hydrogen remain within the supernova opening angle
(perhaps within 30$^{\circ}$ of the rotation axis).

\begin{figure}[htbp]
\plotone{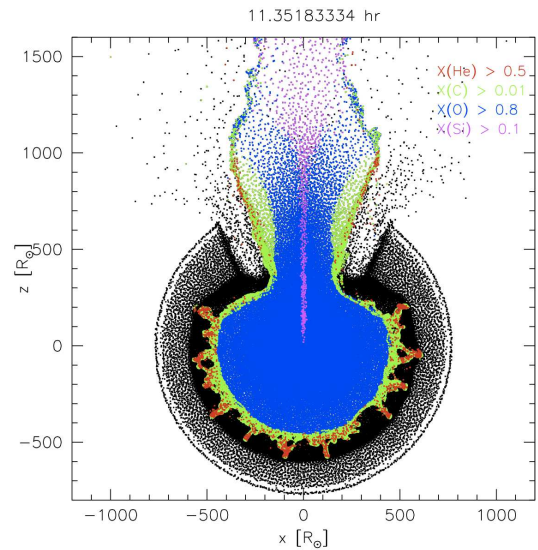}
  \caption{SNSPH particle positions for a $5.6\times10^{51}$\,erg
    explosion focused in a 30$^{\circ}$ wedge of a 25\,M$_\odot$.  The
    density in this wedge is lowered by a factor of 100 with respect
    to the rest of the star.  Shading shows the different chemical
    abundances with black denoting hydrogen and lighter shades showing
    helium, carbon, oxygen and silicon abundances.}
  \label{fig:sn}
\end{figure}

To determine how well the low density funnel can focus the explosion,
we have simulated using the SNSPH code~\citep{FRW06,Ell12} a
$5.6\times10^{51}$\,erg explosion focused in a 30$^{\circ}$ wedge of a
25\,M$_\odot$ where the density in this wedge is lowered by a factor
of 100.  The explosion at 0.47d is shown in figure~\ref{fig:sn}.  In
this calculation, roughly 0.0001\,M$_\odot$ of hydrogen is swept up in
the explosion.  This hydrogen mass is 3 orders of magnitude lower than
typical type IIb supernovae and can easily be hidden in the spectra.  
This supernova would produce a Ib/c supernova.

How much hydrogen is swept up (and whether or not we'll see it in 
the explosion) depends upon the details of the density profile 
set by the merger ejecta and the asymmetry in the explosion.  
For GRB-like jets expected from massive helium-core mergers, 
the amount swept up should not be detectable.  But for main 
sequence stars, the explosion may not be so jet-like, and 
the amount of hydrogen in the explosion could be large.

Finally, like with most collapsar progenitors, the helium in the 
ejecta must either be burned to C/O in the shock, or must be hidden 
in the supernova spectra.  This remains one of the primary open 
problems with the Collapsar GRB model as well~\citep{fryer2007a}.

\section{Comparison to Transient Populations}

We have shown that the rates of our massive he-core explosions are on
par with the long-duration GRB rate, and the total he-merger and 
main-sequence merger rate is roughly 1\% the supernova rate.  For 
massive ($>4\,$M$_\odot$) helium cores, energies can be as high 
as any black-hole accretion disk model.  These models produce a 
range of hydrogen shells surrounding the star at 1-100\,A.U.  But 
to truly understand the nature of the explosions of these mergers 
requires a better understanding of the black-hole accretion disk 
engine.  In this section, we review some of the possible outcomes 
of these objects.

\subsection{He-Mergers and GRBs}

We have shown that for a subclass of merging systems, the accretion
rates onto the black hole can provide the energetics needed to drive a
GRB outburst.  In addition, the rate of this subclass is sufficient to
explain GRBs.  But can these mergers explain the full suite of GRB
diagnostics.  

\cite{Tho11} argued these mergers could fit a very peculiar subset of
GRBs with an unusual optical counterpart.  But can helium mergers
explain a broader set of GRBs?  One of the key constraints of
long-duration GRB progenitors is the association of a particular
subtype of supernova (Type Ic) with GRBs.  The $^{56}$Ni estimates for
these supernovae tend to be higher than typical Ib/c supernovae with
broader line profiles suggesting asymmetric explosions.  Is it
possible for the He-merger progenitor to reproduce this supernova/GRB
association?  

If the helium merger model can not satisfy these requirments, it may 
be limited to explaining only a few odd GRBs such as the Christmas 
burst.  The shell position in the Christmas Burst ($\sim$10\,A.U.) 
is characteristic of our he-merger predictions, but we expect a range 
shell positions from 1-100\,A.U.

One feature of the helium merger model is that it typically has an
order of magnitude higher angular momenta than the collapsar model.
This leads to larger accretion disks with larger accretion times, 
although for the typical specific angular momenta of $10^{18}\, {\rm
  cm^2 s^{-1}}$, the accretion disk remains with $10^{9}$\,cm and may 
not produce drastically different characteristics than the Collapsar 
model.  At this point, our lack of understanding of the GRB engine 
makes it difficult to tie this difference in disk size to an observation.  
But as we better understand the engine, this difference may lead to 
observable consequences.

\subsection{Peculiar Supernovae}

If the explosion is asymmetric but there is too much hydrogen within
the supernova explosion to produce a relativistic jet, helium helium
mergers may produce asymmetric Ib/c supernovae or, if they sweep up
more hydrogen, some fraction of the type IIb supernova population.
These supernovae would be characterized by large asymmetries and,
perhaps, larger $^{56}$Ni yields.

Alternatively, if the disk explosion model produces more symmetric
explosions, the helium merger model is more likely to produce a small
subset of the type IIP supernovae and unless the $^{56}$Ni yield is 
different, it may be difficult to uncover differences between he-merger 
outbursts from typical core-collapse explosions.  Again, it is likely 
that these bursts exhibit stronger asymmetries than normal supernovae and 
evidence of these asymmetries may be the only other characteristic 
differentiating these explosions from normal type IIP supernovae.

Supernovae have been observed that may exhibit a recent shell ejection 
caused by a helium shell ejection (SN2011ht:  Roming et al. 2012).  Using the 
same analysis as section~\ref{sec:hemerger}, we calculate the 
outer shell radii for both our low-mass helium cores and hydrogen 
stars~\ref{fig:rhydist}.  Note that the distribution has many more 
shells with outer radii below 1-2\,A.U.  The explosions from these 
mergers may, at peak, look very similar to comparable giant stars.  
A fraction (6\% of hydrogen stars, 34\% of low-mass helium cores) 
have outer shell radii beyond 10\,A.U. that may produce the peculiar 
features of supernovae such as SN 2011ht.  The broad range of shell 
separations from these systems is in large part due to the fact 
that a sizable fraction of these systems (30\% of hydrogen stars, 
17\% of low-mass helium stars) arise from systems where the 
compact remnant is kicked into its companion. 

\subsection{Radio Bursts}

If, in a GRB-like explosion, the jet sweeps up too much hydrogen to
become relativistic, the engine does not produce much $^{56}$Ni and
supernova explosion is asymmetric (preventing it from sweeping up much
mass), these explosions may produce optical weak, radio loud
outbursts.

\acknowledgements This project was funded in part under the auspices
of the U.S. Dept. of Energy, and supported by its contract
W-7405-ENG-36 to Los Alamos National Laboratory.  K.B.  acknowledges
support from MSHE grant N N203 404939.

{}

\end{document}